\pdfoutput=1
\documentclass[aps,prab,twocolumn,superscriptaddress,longbibliography,floatfix]{revtex4-2}

\usepackage{amsmath,amssymb}
\usepackage{array}[=2016-10-06]
\usepackage{longtable}
\usepackage{graphicx}
\usepackage{xcolor}
\usepackage{bm}
\usepackage[colorlinks=true,linkcolor=blue,citecolor=blue,urlcolor=blue]{hyperref}
\graphicspath{{figs/}}

\newcommand{\wc}{\omega_c}
\renewcommand{\wp}{\omega_p}
\newcommand{\lp}{\lambda_p}
\newcommand{\lc}{\lambda_c}
\newcommand{\n}[1]{n_{#1}}
\newcommand{\Om}{\Omega}
\newcommand{\rcap}{r_{\mathrm{cap}}}

\begin{document}

\title{Magnetizing nonlinear plasma wakefields for positron acceleration: mechanism and operating limits}

\author{Yung-Kun Liu}
\email[Corresponding author: ]{r06222017@ntu.edu.tw}
\affiliation{Leung Center for Cosmology and Particle Astrophysics (LeCosPA), National Taiwan University, Taipei 10617, Taiwan}
\affiliation{Department of Physics, National Taiwan University, Taipei 10617, Taiwan}
\author{Pisin Chen}
\affiliation{Leung Center for Cosmology and Particle Astrophysics (LeCosPA), National Taiwan University, Taipei 10617, Taiwan}
\affiliation{Department of Physics, National Taiwan University, Taipei 10617, Taiwan}
\author{Ching-En Lin}
\affiliation{Department of Applied Physics, Stanford University, Stanford, California 94305, USA}
\affiliation{SLAC National Accelerator Laboratory, Menlo Park, California 94025, USA}
\author{Spencer Gessner}
\affiliation{SLAC National Accelerator Laboratory, Menlo Park, California 94025, USA}
\author{Bernhard Hidding}
\affiliation{Department of Physics, Heinrich Heine University, D\"{u}sseldorf, Germany}

\begin{abstract}
Positron acceleration is possible in linear plasma wakes, but the nonlinear electron-driven blowout regime confines simultaneous acceleration and focusing to narrow regions of returning plasma electrons. Here we investigate how an axial magnetic field reorganizes this nonlinear return to open an operating regime for sustained positron acceleration in a uniform plasma. Using quasi-3D and full-3D particle-in-cell simulations, we connect the formation of finite-radius electron columns, or gyro images, to witness transport and its operating limits. The image spacing is governed by the quasi-static invariant, while canonical angular momentum limits the return radius. Modified by the orbit-sampled wake potential, this mechanism yields a positioning law that accurately predicts the density peaks across a wide range of magnetic field strengths. At a plasma density of $10^{16}\;\mathrm{cm^{-3}}$ and a $35$~T axial field, the reorganized structure expands the usable accelerating and focusing interval by more than a factor of four compared to the unmagnetized case. A positron witness bunch placed at the settled phase of this band achieves $92\%$ capture over a $60$~mm stage, with the normalized emittance saturating at $83$~mm\,mrad. Independent codes confirm sustained transport, yielding energy gains between $99$ and $150$~MeV, though numerical convergence on the exact energy gain remains an open challenge. Together with beam loading, alignment, and driver-energy scaling tests, these results establish magnetic control of the nonlinear electron return as a route to single-stage positron acceleration. Beam quality preservation, formation-transient control, and solenoid integration define the necessary next development steps.
\end{abstract}

\maketitle

\section{Introduction}
\label{sec:intro}

Plasma wakefield accelerators sustain large accelerating gradients~\cite{Tajima1979,Chen1985}, with demonstrated capabilities including electron energy doubling~\cite{Blumenfeld2007}, efficient energy transfer~\cite{Litos2014}, and normalized-emittance preservation~\cite{Lindstrom2024}. Extending these capabilities to positrons is a central challenge for plasma-based $e^+e^-$ colliders~\cite{Cao2024,Joshi2025,ChenLiu2026}. The difficulty is not the absence of a positron-accelerating phase in linear wakes, but rather the access to the nonlinear blowout regime~\cite{Rosenzweig1991,Lu2006}, where the ion cavity inherently focuses electrons and defocuses positrons. In a conventional electron-driven blowout wake, simultaneous positron acceleration and focusing are consequently restricted to short regions where the expelled plasma electrons return toward the propagation axis~\cite{Lotov2007}.

Proposed remedies typically modify the plasma, the driver, or the witness. Hollow channels remove the on-axis transverse force but require external focusing and are susceptible to misalignment-driven wakefields~\cite{Gessner2016,Lindstrom2018}. Thin, warm channels can develop internal focusing structures~\cite{Silva2021}. Finite-radius plasma columns~\cite{Diederichs2019,Diederichs2022} and self-loading positron bunches~\cite{Corde2015,Zhou2025} sustain an electron population near the axis. Hollow electron drivers~\cite{Jain2015} and orbital-angular-momentum laser pulses~\cite{Vieira2014} reshape the sheath, while operation closer to the quasi-linear regime trades accelerating gradient for field regularity~\cite{Doche2017}. These approaches differ in beam requirements, plasma preparation, and attainable beam quality. A brief comparison with the present scheme is provided in Sec.~\ref{sec:compare}.

In this work, we utilize an external axial magnetic field to reorganize the nonlinear electron return, seeking a positron-accelerating regime without reducing the driver to the quasi-linear limit. Magnetic fields have previously been used to guide positrons~\cite{Xu2020} and stabilize drive bunches~\cite{Su1987}, and the linear response of magnetized plasmas to charges and beams has been documented~\cite{Galyamin2013,Galyamin2021}. A recent linearized Green-function treatment reports enhanced focusing and an additional radial eigenmode~\cite{MolaviChoobini2026}, though it does not address the blowout sheath or positron transport. Nonlinear magnetized wakes have also been studied in reduced dimensions~\cite{Nersisyan2000,Balakirev2001,Karmakar2017}. Because the effect considered here relies on transverse electron motion, a strictly one-dimensional wake is unaffected by an axial field~\cite{Nersisyan2000}. Additional contexts include transverse-field control of particle phase~\cite{Katsouleas1983} and magnetized wakes in pulsar plasmas and relativistic astrophysical outflows~\cite{Mofiz1989,ChenTajimaTakahashi2002,Chang2009}.

In laser-driven wakes, the conservation of canonical angular momentum (CAM) can prevent sheath electrons from reaching the axis, thereby inhibiting cavity closure~\cite{Bulanov2013} and reducing trapping~\cite{Rassou2015,Zhao2019}. This same restriction on electron return can alternatively benefit positrons by distributing the returning charge over a finite radius. This dynamic motivated our preliminary two-dimensional proposal for axial-field regularization of an electron-driven wake~\cite{ChenLiu2026}. The accompanying Letter~\cite{Letter} demonstrates positron transport within this structure. The present paper develops the underlying mechanism, the positioning law, numerical validation, and the operating limits.

The magnetic field simultaneously re-times the electron return and limits how closely individual electrons approach the axis, producing successive finite-radius columns that we term gyro images (Sec.~\ref{sec:theory}). This configuration uses a conventional electron driver in a uniform plasma, with a solenoid as the additional hardware element. The unmagnetized wake already possesses accelerating and focusing regions with high local gradients (Sec.~\ref{sec:wake}); however, imposing the magnetic field significantly expands the extent, uniformity, and persistence of this usable wake phase. Sections~\ref{sec:theory} through \ref{sec:tracer} establish this formation mechanism and wake structure, while Secs.~\ref{sec:witness} through \ref{sec:compare} examine transport, loading, tolerances, and scaling.

\begin{figure*}
\includegraphics[width=\textwidth]{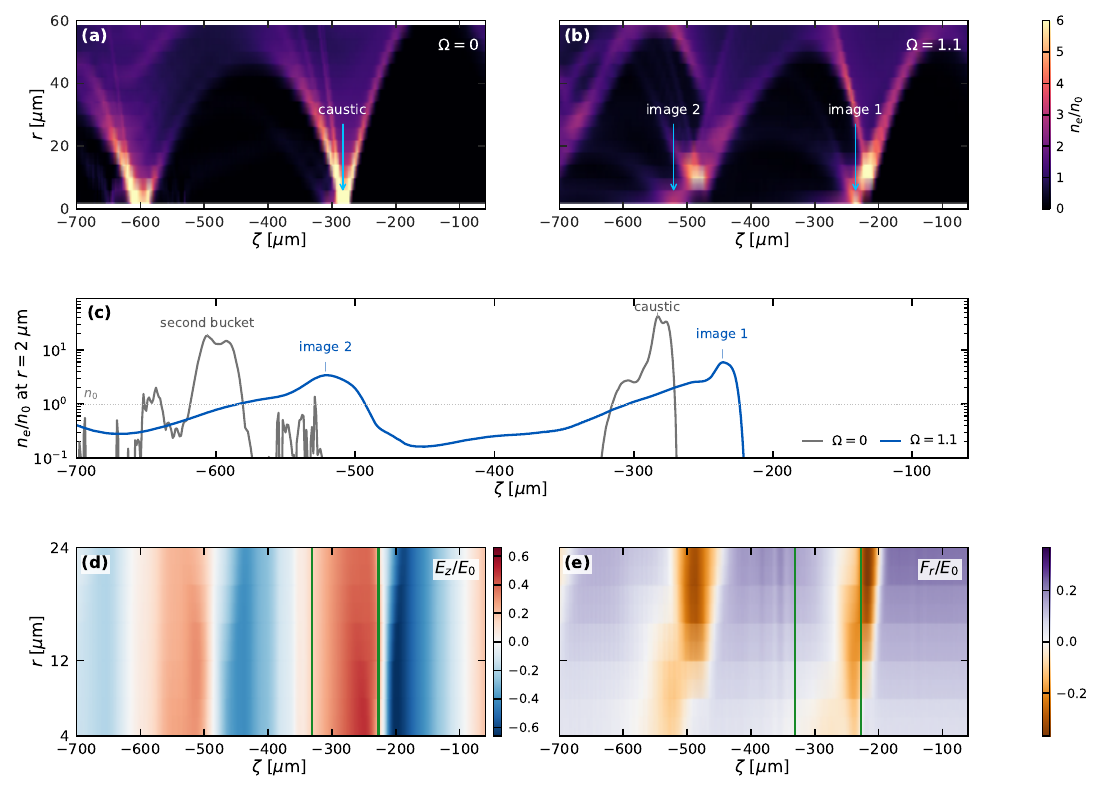}
\caption{\label{fig:column}Returning electron density and positron wakefields at $t=64.7$~ps ($z=19.4$~mm).
(a, b) Density maps behind the driver using the same spatial windows and color scales.
(c) Axial density profiles at $r=2\;\mu$m: the unmagnetized caustic reaches $42\,\n0$, while the magnetized wake forms successive gyro images.
(d, e) Magnetized accelerating field $E_z$ and radial force $F_r=E_r-cB_\theta$, with $11\;\mu$m longitudinal smoothing.
Green lines bound the longest usable interval ($104\;\mu$m) at an acceptance radius of $4\;\mu$m.
The density peak at $-236\;\mu$m differs from the time-averaged position used in the positioning law because the wake migrates (Sec.~\ref{sec:law:bridge}).}
\end{figure*}

Figure~\ref{fig:column} shows how magnetization replaces a narrow density caustic with a broader electron return structure.
At the time shown, the peak density is $42\,\n0$ without the magnetic field and $5.9\,\n0$ at $\Om\equiv\omega_c/\omega_p=1.1$.
For positron acceleration, the crucial consequence is the overlap of accelerating and focusing fields in panels~(d) and (e), whose radial acceptance and transport are examined in Secs.~\ref{sec:wake} and \ref{sec:witness}.

\section{Theory: the gyro clock, the CAM barrier, and the aberration}
\label{sec:theory}

The extended return structure requires both longitudinal organization and a finite transverse scale. We first derive the gyro clock that dictates the longitudinal spacing of successive returns, followed by the angular-momentum constraint that governs their transverse radii. Finally, we account for the wake potentials sampled by realistic orbits.

\subsection{The invariant and the gyro clock}
\label{sec:theory:invariant}
\label{sec:theory:clock}

We adopt the following normalized units throughout this work: potentials are normalized to $m_ec^2/e$, momenta to $m_ec$, lengths to $c/\wp$, times to $\wp^{-1}$, densities to $\n0$, and electric fields to the wave-breaking field $E_0=m_ec\wp/e$. The co-moving coordinate is defined as $\zeta=z-ct$, such that material behind the driver is located at $\zeta<0$.

In the quasi-static approximation, a wake whose potentials depend only on $(\zeta,r)$ conserves the following quantity for electrons initially at rest ahead of the driver~\cite{Mora1997}:
\begin{equation}
\gamma-p_z=1+\psi(\zeta,r),\qquad \psi\equiv\phi-a_z ,
\label{eq:invariant}
\end{equation}
where the pseudo-potential $\psi\to0$ in the unperturbed plasma. A uniform axial magnetic field introduces a vector potential component $A_\theta=B_zr/2$ that is independent of $\zeta$. This external field alters the transverse dynamics without entering the definition of $\phi-a_z$, thereby preserving the invariant in Eq.~(\ref{eq:invariant}).

Consider electrons expelled near a common phase $\zeta_c$ into the quasi-neutral periphery outside the ion cavity. In the ideal limit of negligible wake fields, both $\gamma$ and $p_z$ remain constant, and the transverse momentum $\bm p_\perp$ rotates at the relativistic cyclotron frequency $\Om/\gamma$. One gyro-period lasts
\begin{equation}
T=\frac{2\pi\gamma}{\Om},
\label{eq:period}
\end{equation}
which inherently grows with particle energy. Written as a positive magnitude, the backward co-moving slippage over one such period is
\begin{equation}
\Delta\zeta=(1-\beta_z)\,cT=\frac{2\pi}{\Om}(\gamma-p_z)
=\frac{\lp}{\Om}\bigl(1+\psi\bigr).
\label{eq:imaging}
\end{equation}
Equation~(\ref{eq:invariant}) cancels the explicit $\gamma$ dependence in the slippage. Consequently, at a fixed $\psi$, the co-moving slippage is energy independent even though the individual gyro-period is not. In the limit $\psi\to0$, the spacing interval simplifies to $\lc\equiv\lp/\Om$, yielding return phases at
\begin{equation}
\zeta_N=\zeta_c-N\lc,\qquad N=1,2,\dots
\label{eq:zone}
\end{equation}
These successive returns form the first, second, and subsequent gyro images. We refer to this spacing relation as the gyro clock. Because this relation governs the plasma electron return, it predicts the locations of the density images rather than the evolving usable band or the final witness phase (see Sec.~\ref{sec:law:bridge}). Real orbits inevitably sample a non-zero $\psi$, which introduces an indirect energy dependence through the orbit-dependent potential. Section~\ref{sec:tracer} isolates this potential correction from the relativistic energy dependence of the return time.

\subsection{The CAM barrier sets the transverse scale}
\label{sec:theory:cam}

In the axisymmetric model with a uniform axial field, a second invariant dictates the transverse return scale: the electron canonical angular momentum,
\begin{equation}
\ell=p_\theta r-\frac{\Om r^2}{2}.
\label{eq:cam}
\end{equation}
An electron initially at rest at $r_0>0$ carries an angular momentum magnitude $|\ell|=\Om r_0^2/2$ and cannot reach the axis with a finite momentum. For a small turning radius where the magnetic term in Eq.~(\ref{eq:cam}) becomes negligible compared to $|\ell|$, the minimum approach radius is bounded by
\begin{equation}
r_{\min}\approx\frac{|\ell|}{p_\perp}=\frac{\Om r_0^2}{2p_\perp}.
\label{eq:rmin}
\end{equation}
At a given transverse momentum $p_\perp$, this magnetic barrier is weaker for smaller launch radii, favoring their contribution near the axis. This magnetic exclusion suppresses on-axis density spikes and self-injection in magnetized laser wakefields~\cite{Bulanov2013,Rassou2015,Zhao2019}. While it constrains the return radius, it does not independently determine the collective density profile or guarantee positron focusing. Those structural properties are evaluated numerically in Secs.~\ref{sec:wake} and \ref{sec:witness}.

The transverse dynamics conceptually mirror solenoidal charged-particle optics~\cite{Busch1926,Reiser2008}, while the quasi-static invariant links these dynamics to the co-moving image positions.
At fixed transverse momentum, the gyroradius $r_L=p_\perp/\Om$ and the ideal gyro-image spacing in Eq.~(\ref{eq:zone}) diverge as $\Om\to0$.
These scales describe gyro motion with negligible wake fields, as assumed in deriving the clock; the wake fields can still drive electron returns on a finite timescale.
In the unmagnetized limit, the CAM barrier in Eq.~(\ref{eq:cam}) also vanishes, allowing returning electrons to reach the axis and form a density cusp.

\subsection{Aberration and range of validity}
\label{sec:theory:aberration}
\label{sec:theory:validity}

When electrons sample a nonuniform $\psi$, the gyro-clock approximation depends on the gyrophase-weighted orbit mean $\bar\psi$, where $\mathrm{d}\alpha\propto\mathrm{d}t/\gamma$. The ensemble mean shifts the image spacing through an effective wavelength $\lambda_{p,\mathrm{eff}}\simeq\lp(1+\bar\psi)$, while the orbit-to-orbit potential spread introduces a leading-order longitudinal blur:
\begin{equation}
\delta\zeta_N\simeq N\,\delta\bar\psi\,\frac{\lp}{\Om}.
\label{eq:blur}
\end{equation}
Assuming comparable potential spreads, the second image is therefore approximately twice as broad as the first. The dependence on launch radius enters through the specific potential history sampled by each orbit; recalibrating the mean wavelength cannot eliminate this inherent spread. Section~\ref{sec:tracer:closure} compares the field-sampled $\bar\psi=+0.075$ with the independently calibrated spacing to verify consistency.

The gyro-clock model relies on a well-defined electron expulsion phase and quasi-static fields. Empirically, this requires a sufficiently strong driver ($\phi_0\equiv n_b/\n0\gtrsim2$); for weaker drivers, the positioning law degrades because the expulsion phase is no longer sharply defined. Furthermore, because real electrons travel through non-zero wake potentials, their actual co-moving slippage deviates from the ideal field-free limit. This deviation necessitates the orbit-averaged correction introduced above to accurately predict the image spacing.

For $\Om\sim1$, the nominal cyclotron and plasma frequencies are comparable, and the magnetic field substantially modifies the electron return motion.
The gyro clock describes the phase organization acquired in the quasi-neutral periphery, while the self-consistent wake fields modify the orbits and shift the image phases.
A density image is useful for positron acceleration where its fields provide simultaneous acceleration and focusing over a finite radius.
Section~\ref{sec:wake:scan} identifies this operating window from the field overlap.

\section{Numerical methods and validation protocol}
\label{sec:methods}

To evaluate the return mechanism as a viable accelerating structure, we utilize a local field criterion to identify candidate operating regions and particle diagnostics to test beam transport within them. These two tests remain distinct; their numerical sensitivities are examined in Sec.~\ref{sec:methods:robustness} and the subsequent transport comparisons.

\subsection{Simulation setup}
\label{sec:methods:setup}

The baseline ($\Om=1.1$) and full-3D calculations utilize the particle-in-cell code Smilei~\cite{Derouillat2018}. The cross-code and fine-resolution transport benchmarks use WarpX~\cite{Vay2018} (Sec.~\ref{sec:witness:convergence}). The baseline applies an azimuthal Fourier decomposition in cylindrical geometry~\cite{Lifschitz2009,Derouillat2018,Zemzemi2020}, retaining modes $m\le1$. Stability runs extend this truncation to $m\le3$ (Sec.~\ref{sec:stability}), and the full-3D Cartesian runs operate without azimuthal mode truncation. Table~\ref{tab:setup} lists the numerical parameters, while Table~\ref{tab:runs} (Appendix~\ref{app:impl:runs}) inventories the specific witness-transport runs.

\begin{table*}
\caption{\label{tab:setup}Numerical parameters for the quasi-3D and full-3D simulations. The $60$~mm witness transport run and the $\Om$-scan runs share the quasi-3D parameters. Here, $m$ denotes the azimuthal Fourier mode number, FDTD stands for finite-difference time-domain, and CFL refers to the Courant limit of the time step.}
\begin{ruledtabular}
\begin{tabular}{lll}
 & quasi-3D (baseline) & full-3D \\
\hline
geometry & quasi-cylindrical (\texttt{AMcylindrical}), $m\le1$ & Cartesian \\
cells & $2048\times256$ ($z\times r$) & $512\times96\times96$ \\
cell size & $\Delta z=1\;\mu$m, $\Delta r=4\;\mu$m &
  $\Delta x=1\;\mu$m, $\Delta y=\Delta z=4\;\mu$m \\
box & $2048\times1024\;\mu$m & $512\times384\times384\;\mu$m \\
time step & $0.9\,\Delta z/c$ ($0.75$ in $m\le3$ runs) &
  $0.8\,\Delta t_{\rm CFL}=0.754\,\Delta x/c$ \\
duration & $200$~ps ($60$~mm); $65$~ps (scan) &
  $200$~ps ($60$~mm); $100$~ps (two placement runs) \\
particles/cell & 32 ($e^-$) / 8 (ion) / 32 (driver) & 8 / 1 / 8 \\
ions & mobile H$^+$, $m_i/m_e=1836$ & same \\
solver & Yee FDTD; Silver--M\"uller ($z$), Buneman ($r$) &
  Yee FDTD; Silver--M\"uller (all faces) \\
interpolation & order 2 & order 2 \\
driver init & relativistic Poisson solver & same \\
axial field & \texttt{ExternalField} $B_{z,m=0}=\Om$ &
  prescribed analytic $B_x=\Om$ \\
moving window & at $c$, from $t=4.1$~ps & at $c$, from $t=0$ \\
\end{tabular}
\end{ruledtabular}
\end{table*}

The simulation parameters are normalized to a uniform plasma density $\n0=10^{16}\,\mathrm{cm^{-3}}$. This sets the length scale $c/\wp=53.15\;\mu$m, the wavelength $\lp=334\;\mu$m, the reference field $E_0=9.6$~GV/m, and the timescale $\wp^{-1}=0.177$~ps. The magnetic-field unit translates to $m_e\wp/e=32.1$~T per unit $\Om$; therefore, the $\Om=1.1$ baseline corresponds to $35.3$~T, referenced as $35$~T hereafter.

The driver is initialized via a relativistic Poisson solver using the space-charge fields of a $\gamma_d=1000$ electron beam. The beam possesses a Gaussian profile truncated at the $4\sigma$ boundary, with $1/e$ half-widths $w_z=15.9\;\mu$m and $w_r=50\;\mu$m, a peak density $n_b/\n0=2.5$, and a total charge of $0.889$~nC ($0.454$~J). The driver is initialized as dynamically cold, possessing zero emittance, zero energy spread, and a uniform $\beta_z$. It enters a sharp-edged plasma without an entrance density ramp. In the full-3D runs, the propagation axis is the Cartesian $x$-axis (corresponding to the $z$-axis in quasi-3D runs), and the uniform axial magnetic field is applied as a prescribed analytic field (Appendix~\ref{app:impl}).

The Gaussian positron witness consists of $2\times10^4$ macroparticles ($4\times10^5$ in the full-3D runs), with $\sigma_r=7\;\mu$m, $\sigma_z=4\;\mu$m, $\gamma_w=5000$, and a normalized emittance $\varepsilon_n=50$~mm\,mrad. The total witness charge is $0.1$~pC, corresponding to a peak density of $0.02\,\n0$, serving as a weak-loading probe. Heavier beam loading is examined separately in Sec.~\ref{sec:witness:loading}. The $60$~mm run simultaneously carries two identical witnesses: a main bunch at $\zeta=-227\;\mu$m and a control bunch at $\zeta=-245\;\mu$m.

Appendix~\ref{app:impl} specifies further numerical details, including the cylindrical-axis ghost row, the staggered field grids, and the mode-dependent Courant limit.

\subsection{The usability criterion}
\label{sec:methods:criterion}

The field criterion identifies regions providing simultaneous acceleration and focusing over a finite radial extent. Following the convention in Sec.~\ref{sec:theory:invariant}, $E_z>0$ accelerates and $F_r<0$ focuses a forward-moving ultrarelativistic positron. For an acceptance radius $R$, we classify an interval in $\zeta$ as usable when it satisfies three conditions:
\begin{enumerate}
\item[(i)] The near-axis accelerating field satisfies $E_z(4\;\mu\mathrm{m})>0.10\,E_0$.
\item[(ii)] The radial wake force on an ultrarelativistic, paraxial positron ($F_r\equiv E_r-B_\theta$) is focusing ($F_r<0$) at every sampled radius within $4\;\mu\mathrm{m}\le r\le R$. Transverse-velocity terms involving $B_z$ are omitted in this static field metric, though the witness runs in Sec.~\ref{sec:witness} naturally integrate the full Lorentz force.
\item[(iii)] The accelerating field $E_z$ at those radii remains within $20\%$ of its value at $4\;\mu$m.
\end{enumerate}

The longest continuous interval satisfying these conditions is defined as the \emph{usable band}, with length $L(R)$. The aperture sweep evaluates $L(R)$ by varying $R$ and reapplying the clauses (Sec.~\ref{sec:wake:lofr}). The fields are evaluated individually per snapshot rather than time-averaged, as temporal averaging can mathematically produce an interval that does not physically exist at any given moment. Prior to evaluation, the fields are longitudinally smoothed over $11\;\mu$m, which is short compared to the image spacing ($\lambda_{p,\mathrm{eff}}/\Om=328\;\mu$m). Positions within $60\;\mu$m behind the driver are excluded. The reference radius $4\;\mu$m is held constant across resolutions.

For the field-strength scan presented in Sec.~\ref{sec:wake:scan}, we modify the approach to use time-averaged maps and demand acceleration and focusing specifically at a single scoring radius, $r=8\;\mu$m, alongside radial uniformity between $r=4$ and $24\;\mu$m (Appendix~\ref{app:sens}). The longest qualifying interval under these rules is termed the \emph{quality interval}. Because both the temporal treatment and radial sampling differ, the length of the quality interval is distinct from $L(R)$.

While these local criteria identify candidate regions, they are not strictly necessary for particle confinement; a fully resolved settled wake can transport a witness despite a weak-force core (Sec.~\ref{sec:witness:fine}).

\subsection{Witness diagnostics}
\label{sec:methods:diagnostics}

Witness statistics are charge-weighted over the \emph{captured ensemble}, defined at each output as the particles inside the capture radius $\rcap=21.5\;\mu$m.
The capture fraction describes charge retention, while the energy gain, spread, and emittance characterize the bunch carried by the wake.
Diagnostic cadence and the effect of ensemble selection are given in Appendix~\ref{app:impl:witness}.

\section{Wake structure, the $\Om$ scan, and the life cycle of the band}
\label{sec:wake}

The returning electron flux shown in Fig.~\ref{fig:column} produces a usable band whose spatial extent depends on the acceptance radius, the magnetic field strength, and the propagation distance. To analyze its evolution, we distinguish three temporal phases: the \emph{scan window} comprising four snapshots at $43.1$--$64.7$~ps ($13$--$19$~mm); the subsequent \emph{breathing} excursion extending to approximately $36$~mm; and the \emph{settled era} ($z\ge44$~mm) where the baseline band edges remain approximately stationary.

\subsection{Usable length and its aperture dependence}
\label{sec:wake:lofr}

\begin{figure}
\includegraphics[width=\columnwidth]{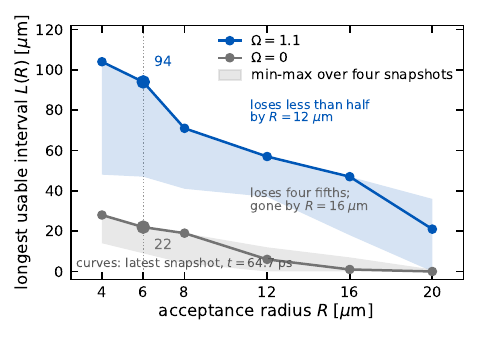}
\caption{\label{fig:lofr}Usable length $L(R)$ as a function of the acceptance radius $R$, evaluated in each configuration's optimal region. Solid curves display $L(R)$ at $t=64.7$~ps ($z=19.4$~mm), while the shaded regions indicate the variation across four snapshots spanning $43.1$ to $64.7$~ps. Larger filled circles highlight $R=6\;\mu$m, the smallest tested radius where the radial uniformity constraint actively limits the usable length.}
\end{figure}

Applying the criteria from Sec.~\ref{sec:methods:criterion} to both the unmagnetized and magnetized configurations at identical propagation distances yields the aperture sweeps shown in Fig.~\ref{fig:lofr}. Each configuration is evaluated in its optimal region, including the second bucket.

At $t=64.7$~ps ($19.4$~mm), the magnetized configuration retains significantly more usable length as the acceptance radius $R$ increases. Specifically, for $R=4$, $8$, $12$, $16$, and $20\;\mu$m, the unmagnetized usable lengths $L(R)$ are $28$, $19$, $6$, $1$, and $0\;\mu$m, respectively. In contrast, the magnetized wake provides $104$, $71$, $57$, $47$, and $21\;\mu$m. Furthermore, averaged over this usable island, the radial nonuniformity of $E_z$ between $r=4$ and $12\;\mu$m is $6.3\%$ with the field versus $36.2\%$ without it.

Across the four scan-window snapshots, the longest usable island at $R=4\;\mu$m ranges from $48$ to $104\;\mu$m with the magnetic field, compared to $14$ to $28\;\mu$m without it. This represents a snapshot-by-snapshot contrast of $2.4$ to $3.7$ in usable length and an improvement in radial uniformity by a factor of four to nine. At $R=6\;\mu$m, the magnetized lengths span $47$--$94\;\mu$m against $9$--$22\;\mu$m for the unmagnetized case. Table~\ref{tab:snapshots} details these individual snapshots and the radial sampling procedure.

\subsection{Robustness of the field criterion}
\label{sec:methods:robustness}

To ensure these results are not artifacts of specific threshold choices, we conducted sensitivity tests by varying the accelerating threshold ($0.05$, $0.10$, and $0.15\,E_0$), the radial-uniformity tolerance ($10$, $20$, and $30\%$), and the smoothing kernel ($7$, $11$, and $15\;\mu$m). On the fixed analysis window described in Appendix~\ref{app:sens:criterion}, all 36 combinations confirm a magnetized-to-unmagnetized length contrast of at least a factor of two at $R=4\;\mu$m. At $R=6\;\mu$m, where the uniformity tolerance becomes active, this contrast increases to between $5.2$ and $23.5$.

The length advantage diminishes if the criterion heavily prioritizes gradient over spatial extent. At a higher threshold of $0.40\,E_0$, the unmagnetized wake still supports a short usable band, whereas the magnetized band vanishes on the baseline grid (Appendix~\ref{app:sens:criterion}). At $64.7$~ps, the island-averaged accelerating field is $5.69$~GV/m without the field and $3.27$~GV/m with it. Therefore, the primary benefit of the magnetic field is the spatial extent and radial uniformity of the usable region, rather than a locally higher gradient.

\subsection{Longitudinal flatness}
\label{sec:wake:flatness}

Because radial uniformity does not inherently limit the energy variation along a bunch, we separately examine longitudinal flatness. Table~\ref{tab:flatness} lists the longest approximately flat segments within each island at $64.7$~ps.

At a $\pm3\%$ tolerance, the longest flat segment occurs in the second magnetized image, spanning $22\;\mu$m. The $104\;\mu$m first-image island contains a $19\;\mu$m flat segment, compared with $11\;\mu$m in the unmagnetized wake. This flatness contrast is less pronounced than the usable-length contrast.

\begin{table}
\caption{\label{tab:flatness}Longitudinal flatness island by island, at $t=64.7$~ps.
Each entry is the longest window inside that island over which $E_z(4\;\mu\mathrm{m})$ stays within the stated tolerance of the window midpoint, in micrometers; the island's own length is given with its name. }
\begin{ruledtabular}
\begin{tabular}{lcccc}
island & $\pm3\%$ & $\pm5\%$ & $\pm10\%$ & $E_z$ at mid. \\
\hline
$\Om=1.1$, image 1 ($104$) & $19$ & $25$ & $38$ & $0.469\,E_0$ \\
$\Om=1.1$, image 2 ($79$) & $22$ & $37$ & $53$ & $0.281\,E_0$ \\
$\Om=0$, longest ($28$) & $11$ & $15$ & $24$ & $0.604\,E_0$ \\
$\Om=0$, bucket 2 ($7$) & $1$ & $2$ & $3$ & $0.310\,E_0$ \\
\end{tabular}
\end{ruledtabular}
\end{table}

\subsection{The working window in $\Om$}
\label{sec:wake:scan}

\begin{figure}
\includegraphics[width=\columnwidth]{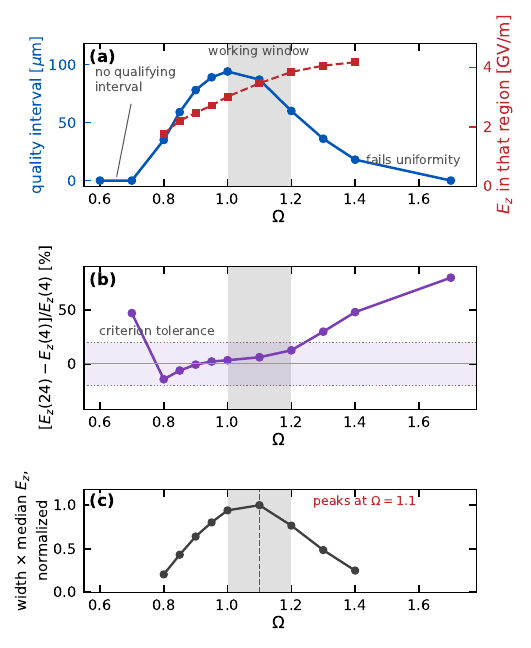}
\caption{\label{fig:window}Magnetic field-strength scan based on the time-averaged quality interval (see Sec.~\ref{sec:methods:criterion}), evaluated with a focusing requirement at $r=8\;\mu$m. (a) Length of the quality interval (blue, left axis) and its median accelerating field (red, right axis). Zero width indicates that no interval meets the quality criterion. (b) Signed radial variation of the accelerating field between $r=4$ and $24\;\mu$m evaluated at the band's optimal phase; shading bounded by dotted lines indicates the $\pm20\%$ tolerance threshold. (c) The product of interval width and median field, normalized to its peak value. The gray vertical band denotes the identified optimal working window of $\Om=1.0$--$1.2$.}
\end{figure}

The magnetic-field scan balances accelerating strength against the extent of the field overlap (Fig.~\ref{fig:window}).
Under the time-averaged quality criterion defined in Sec.~\ref{sec:methods:criterion}, no interval qualifies at the sampled fields $\Om=0.6$ and $0.7$.
The quality interval reaches $94$ and $87\;\mu$m at $\Om=1.0$ and $1.1$, and closes again at $1.7$.
Where it exists, its median accelerating field rises from $1.77$~GV/m at $\Om=0.8$ to $4.18$~GV/m at $1.4$.

The product of interval width and field strength peaks at $\Om=1.1$.
At stronger fields the accelerating field becomes less uniform across the aperture: its signed variation between $r=4$ and $24\;\mu$m exceeds $20\%$ at the band's working point between $\Om=1.2$ and $1.3$.
These trends identify a working window near $\Om=1.0$--$1.2$ ($32$--$38$~T at $\n0=10^{16}\;\mathrm{cm^{-3}}$), with $\Om=1.1$ ($35$~T) used for the baseline.
The dependence on the required focusing aperture is given in Appendix~\ref{app:sens:fields}.

The working window reflects the overlap and radial uniformity of the accelerating and focusing fields.
Density peaks remain identifiable outside this window, including at $\Om=1.7$, where the quality interval has closed.
Thus, the existence of a gyro image and its suitability for transporting a finite-width positron bunch are separate physical requirements.

\subsection{The life cycle of the band}
\label{sec:wake:lifecycle}

\begin{figure}
\includegraphics[width=\columnwidth]{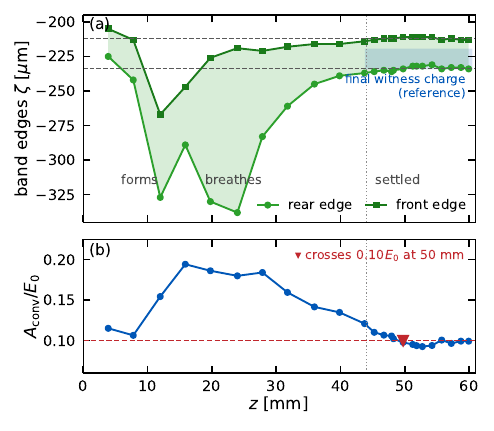}
\caption{\label{fig:lifecycle}Evolution of the usable band and wake amplitude in the baseline $60$~mm run, sampled across $24$ snapshots ($z=3.9$--$59.9$~mm). (a) Dynamics of the $R=4\;\mu$m usable band, showing its rear edge (circles), front edge (squares), and interior extent (green shading). The vertical dotted line marks the onset of the settled era at $z=44$~mm, with horizontal dashed lines indicating the mean band edges thereafter. The blue bar illustrates the final central $95\%$ footprint of the captured witness charge, demonstrating its alignment with the settled band. (b) The near-driver smoothed wake amplitude $A_{\rm conv}$. The dashed red line indicates the $0.10\,E_0$ operational evaluation threshold.}
\end{figure}

Figure~\ref{fig:lifecycle} tracks the baseline usable band at $R=4\;\mu$m.
It forms with a length of $20\;\mu$m at $z=3.9$~mm and broadens to $119\;\mu$m near $24$~mm, with its rear edge at $-338\;\mu$m.
The subsequent contraction is dominated by the forward motion of the rear edge: from $28$~mm onward, this edge marks the boundary of focusing, while the front edge is set by the accelerating-field threshold.
Across fourteen snapshots at $z=43.8$--$59.9$~mm, the edges lie at $-233.9\pm1.7$ and $-212.1\pm1.0\;\mu$m (mean $\pm$ temporal standard deviation), with a median length of $21\;\mu$m.

The field overlap evolves as the driver contracts and subsequently expands.
Increasing the driver energy delays both the pinch and band settling while recovering a similar late-time band (Sec.~\ref{sec:scaling:energy}).
Settling therefore describes a stable accelerating--focusing overlap during continuing driver evolution.
The near-driver amplitude $A_{\rm conv}$ falls from $0.194\,E_0$ at $16$~mm to below $0.10\,E_0$ near $50$~mm, while the usable band and witness acceleration persist to the stage endpoint.

The final central $95\%$ charge footprint of the witness bunch occupies $[-234.4,-219.4]\;\mu$m. This aligns closely with the settled band: the rear tip lies $0.5\;\mu$m beyond the mean rear edge (well within the scatter), and the front sits about seven micrometers inside the band. Transport within this evolving structure is evaluated in Sec.~\ref{sec:witness}.

\section{The positioning law and the $\lambda_{p,\mathrm{eff}}$ calibration}
\label{sec:law}

\begin{figure}
\includegraphics[width=\columnwidth]{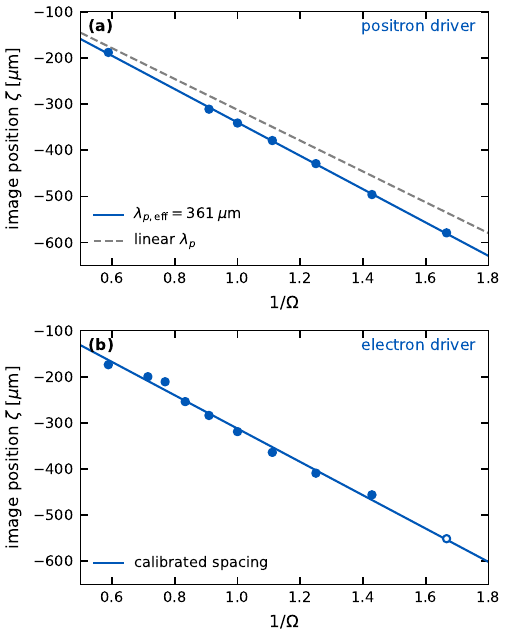}
\caption{\label{fig:law}Gyro-image position versus inverse magnetic field strength.
(a) The positron-driver scan calibrates the effective wavelength to $\lambda_{p,\mathrm{eff}}=361\;\mu$m; the dashed line uses the linear plasma wavelength with the same anchor.
(b) The electron-driver scan follows the same spacing law with a fitted anchor $\zeta_c=+48.6\;\mu$m.
Symbols show the density peaks associated with the first gyro image; the open circle marks the weak feature at $\Om=0.6$.
The peak-selection procedure and residual checks are given in Appendix~\ref{app:sens:selection}.}
\end{figure}

The gyro clock predicts how the electron-return positions move as the magnetic field changes.
We test this relation using the near-axis density peaks formed during the early wake evolution, and then connect these collective images to individual return orbits in Sec.~\ref{sec:tracer}.

\subsection{Calibrating $\lambda_{p,\mathrm{eff}}$ with a positron driver}
\label{sec:law:calibration}

The calibration uses the same plasma and grid parameters with the driver charge reversed.
Across seven field strengths ($\Om=0.6$--$1.7$), fitting the image positions to $\zeta_1=\zeta_c-\lambda_{p,\mathrm{eff}}/\Om$ gives $\lambda_{p,\mathrm{eff}}=361\;\mu$m and a $2.2\;\mu$m rms residual [Fig.~\ref{fig:law}(a)].
The effective wavelength is $8\%$ longer than the linear plasma wavelength $\lambda_p=333.9\;\mu$m, consistent with the wake-potential correction to the gyro clock.
We hold this wavelength fixed when testing the electron-driver scan; the tracer analysis in Sec.~\ref{sec:tracer:closure} provides a separate estimate from the potentials sampled along the orbits.

\subsection{Image positions in the electron-driven wake}
\label{sec:law:fit}

With $\lambda_{p,\mathrm{eff}}=361\;\mu$m fixed, the ten electron-driver runs give a common anchor $\zeta_c=+48.6\;\mu$m and a $9.4\;\mu$m rms residual [Fig.~\ref{fig:law}(b)].
Allowing the wavelength to vary gives $359.7\;\mu$m, within $0.4\%$ of the positron-driver calibration.
In these scans, changing the driver charge preserves the effective spacing while allowing a different expulsion phase.
The peaks are associated with the first gyro image using the positioning law; selecting the most prominent peaks independently retains the $1/\Om$ trend with a $21.5\;\mu$m rms residual.
Appendix~\ref{app:sens:selection} gives the extraction method and sensitivity checks.

\subsection{Driver-profile dependence}
\label{sec:law:profile}

Replacing the Gaussian transverse driver profile with triangular or flat-top profiles (holding total charge and rms size fixed) shifts the effective image spacing by $-2.1\%$ and $-5.0\%$, respectively. Physically, this weak dependence reflects the minor modifications to the orbit-averaged wake potential $\bar{\psi}$ caused by the altered sheath geometry.

The usable field overlap is more sensitive to the driver profile than the image spacing.
For the Gaussian and flat-top drivers, the median centers of the longest $R=4\;\mu$m intervals are $-279$ and $-239.5\;\mu$m over the scan snapshots.
Thus, a similar return spacing can coexist with a substantially shifted working phase.

The flat-top driver also produces a stronger but shorter usable region: the median island-averaged field is $0.46\,E_0$, compared with $0.34\,E_0$ for the Gaussian, while the median length falls from $58$ to $33\;\mu$m.
The corresponding median radial variation rises from $8.4\%$ to $12.6\%$.
These statistics use physical radial coordinates and the same five snapshots at $35.6$--$64.7$~ps, illustrating how driver shaping changes the phase and length--gradient balance of the field overlap.

\subsection{Mapping the initial density peak to the final witness phase}
\label{sec:law:bridge}

\begin{figure}
\includegraphics[width=\columnwidth]{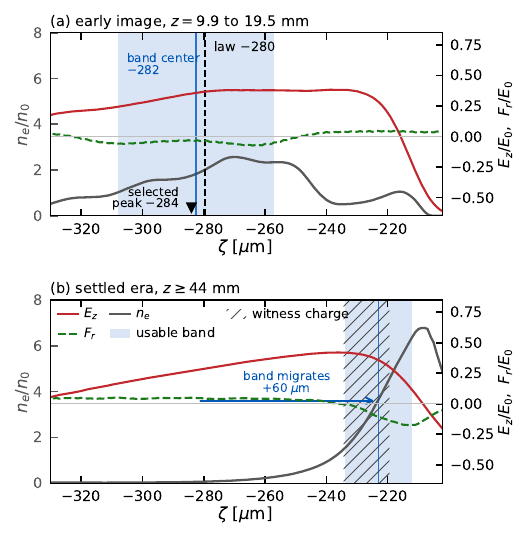}
\caption{\label{fig:bridge}Relationship between the forming density image and the final witness placement in the baseline $60$~mm run. Profiles display temporal medians of electron density at $r=2\;\mu$m (grey, left axis), and of $E_z$ (red) and $F_r$ (green dashed) at $r=4\;\mu$m (right axis); $F_r$ is interpolated from its staggered grid. The blue shaded region spans the mean snapshot-based band edges. (a) The early analysis window ($z=9.9$--$19.5$~mm, two snapshots). The initial band closely aligns with the selected density peak (triangle at $-284\;\mu$m) and the theoretical positioning law (dashed line at $-279.6\;\mu$m). (b) The settled era ($z=43.8$--$59.9$~mm, $14$ snapshots). The usable band has migrated forward by $59.5\;\mu$m to a stable position. The hatched region indicates the footprint of the successfully transported witness bunch.}
\end{figure}

At $\Om=1.1$, the selected initial density peak lies at $\zeta=-284\;\mu$m, whereas the baseline witness is placed at $-227\;\mu$m.
Figure~\ref{fig:bridge} relates these phases through the migration of the accelerating--focusing overlap.

The early band center averages $-282.5\;\mu$m, close to the initial density image, and subsequently moves forward by about $60\;\mu$m to the settled center at $-223.0\;\mu$m.
This shift motivates the empirical baseline placement at $\zeta_w=-227\;\mu$m.
The positioning law locates the forming image; the evolving field overlap determines the phase available to the witness.
Appendix~\ref{app:sens:settling} gives the sampling behind this comparison.

\section{Tracer electrons: connecting return orbits to gyro images}
\label{sec:tracer}

The density images describe the collective return of plasma electrons.
To resolve the motion behind them, we follow $2800$ negligibly loading tracers launched at radii $r_0=2$--$26\;\mu$m and compare their trajectories with an unmagnetized control.
The first perigee is the electron's closest approach to the axis during its first return; its time $\tau^*$ is referenced to the passage of the driver centroid.
The sampling and perigee-extraction procedure are specified in Appendix~\ref{app:sens:tracer}.

\subsection{Return geometry and timing}
\label{sec:tracer:time}
\label{sec:tracer:control}

\begin{figure}
\includegraphics[width=\columnwidth]{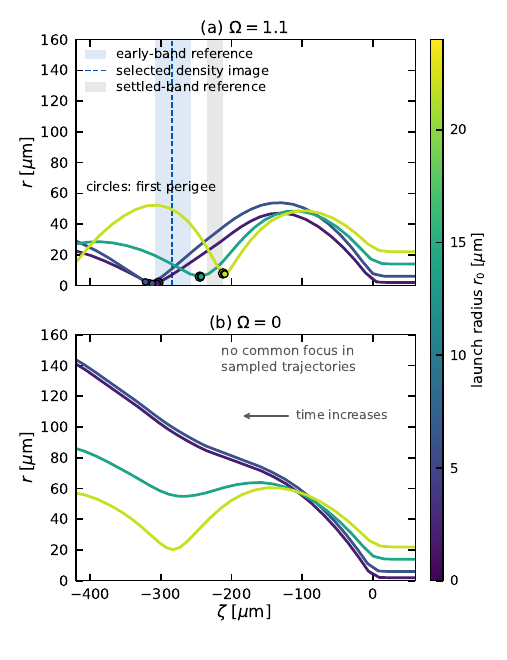}
\caption{\label{fig:orbits}Representative co-moving trajectories sampled every $0.05\,T_c$ during the early formation window ($t=33$--$48$~ps), prior to the settled era. Colors denote launch radii; time increases from right to left. (a) In the magnetized wake ($\Om=1.1$), trajectories converge to form gyro images. Circles mark the first perigee for each track. The blue and grey bands denote the locations of the early and settled usable bands, respectively; the dashed blue line marks the selected density image. (b) In the unmagnetized control ($\Om=0$), the sampled inner trajectories move outward and lack a common return focus.}
\end{figure}

Figure~\ref{fig:orbits} shows how magnetization redirects the inner plasma electrons into a common return region.
In the unmagnetized control, particles launched at $r_0\le14\;\mu$m instead move outward, with median minimum radii of $50$--$57\;\mu$m and no crossing inside the capture radius $\rcap$.
Only $6\%$ of the $r_0=22\;\mu$m bin pass inside $\rcap$, without forming a common gyro image.
The caustic in Fig.~\ref{fig:column} therefore draws on a different part of the electron sheath than these inner launch radii.

The magnetized tracers return at a median time $\tau^*=0.920\,T_c$, where $T_c=2\pi/(\Om\wp)$.
Their transverse momentum rotates through a median of $0.86$ turns before perigee, with the magnetic contribution accounting for $92\%$ of that rotation.
The first approach is therefore a partial orbit, explaining how it can precede $T_c$ even though a free relativistic gyro-period is longer.
Bin-median return times decrease from $1.10\,T_c$ at $r_0=6\;\mu$m to $0.74\,T_c$ at $22\;\mu$m (Table~\ref{tab:tracer}).
The image positions also depend on the longitudinal displacement accumulated during this return.

\subsection{Finite return radius and the CAM barrier}
\label{sec:tracer:depth}

\begin{figure}
\includegraphics[width=\columnwidth]{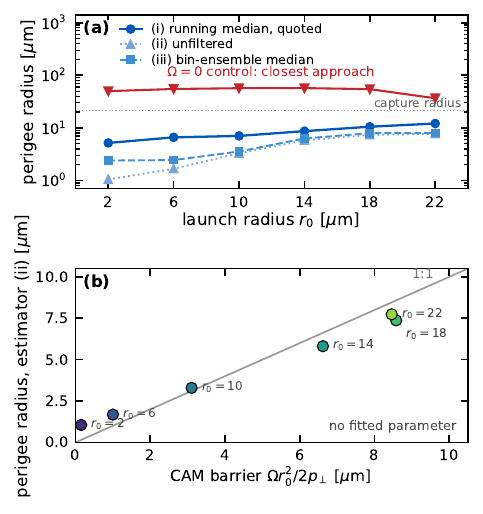}
\caption{\label{fig:perigee}Perigee depth and the canonical angular momentum (CAM) estimate. (a) Median perigee radii for the six launch-radius bins under three different estimators (see Table~\ref{tab:tracer}). Red triangles indicate the bin-median minimum radius for the unmagnetized control run, and the dotted line represents the capture radius. (b) Bin medians of the unfiltered radius at the selected perigee time against bin medians of the theoretical CAM barrier $|\ell|/p_\perp=\Om r_0^2/(2p_\perp)$.}
\end{figure}

Electrons launched closer to the axis reach smaller perigee radii, following the launch-radius dependence of canonical angular momentum (Fig.~\ref{fig:perigee}).
Using each particle's momentum at perigee, the CAM estimate $r_{\min}\simeq|\ell|/p_\perp$ agrees with the unfiltered bin-median radii to within $1.5\;\mu$m in the four outer plotted bins.
The conserved angular momentum thus supplies a transverse scale for the return.
The motion is recurrent: $2762$ of the $2800$ tracers reach a second perigee, at a median time of $1.96\,T_c$.
Its broader distribution in $\zeta$ is consistent with the accumulation of image blur in Eq.~(\ref{eq:blur}).

\subsection{Wake-potential correction and the image position}
\label{sec:tracer:closure}
\label{sec:tracer:anchor}

\begin{figure}
\includegraphics[width=\columnwidth]{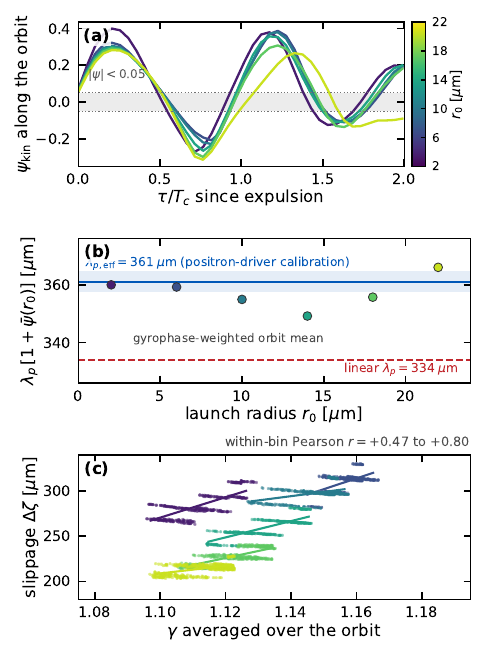}
\caption{\label{fig:tracer}Slippage and potential diagnostics for the tracer ensemble. (a) Bin-median kinematic pseudo-potential $\psi_{\rm kin}=\gamma-p_z-1$ over time; shading denotes the small-potential interval $|\psi|<0.05$. (b) The implied full-turn effective wavelength $\lambda_p[1+\bar\psi(r_0)]$, derived from the bin mean of field-sampled orbit averages weighted by $\mathrm{d}t/\gamma$ (error bars denote standard error across $400$ particles per bin). The solid line marks the positron-driver calibration ($\lambda_{p,\mathrm{eff}}=361\;\mu$m) with $\pm1\%$ shading, while the dashed line indicates the linear plasma wavelength. (c) Co-moving slippage versus time-averaged orbit energy, displaying linear fits within each launch-radius bin (colors match panel a).}
\end{figure}

The potentials sampled along the return orbits account for the effective wavelength exceeding the linear plasma wavelength.
Reconstructing $\psi_{\rm field}$ from the same run's fields supports the quasi-static invariant along the trajectories (Appendix~\ref{app:impl:psi}).
The orbit means, weighted by $\mathrm{d}t/\gamma$, have an ensemble median $\bar\psi=+0.075$, corresponding to $\lambda_p(1+\bar\psi)\simeq359\;\mu$m.
This full-turn estimate is close to the separate positron-driver calibration of $361\;\mu$m [Fig.~\ref{fig:tracer}(b)].
Orbit-dependent potential histories also connect slippage to particle energy [panel (c)] and broaden the collective image.

For the actual partial orbits, the median first-perigee position is $\zeta=-262.5\;\mu$m, ahead of the full-turn interval of $328.2\;\mu$m behind the driver.
The three inner launch bins reach perigee at median positions of $-282$, $-311$, and $-298\;\mu$m.
The innermost bin lies close to the density peak at $-284\;\mu$m and gives an effective anchor of about $+47\;\mu$m, compared with the collective fit of $+48.6\;\mu$m.
The ensemble median mixes different launch radii; the inner returns provide the more direct connection to the near-axis image.

\section{Witness transport, resolution dependence, and cross-code comparison}
\label{sec:witness}

\begin{figure}
\includegraphics[width=\columnwidth]{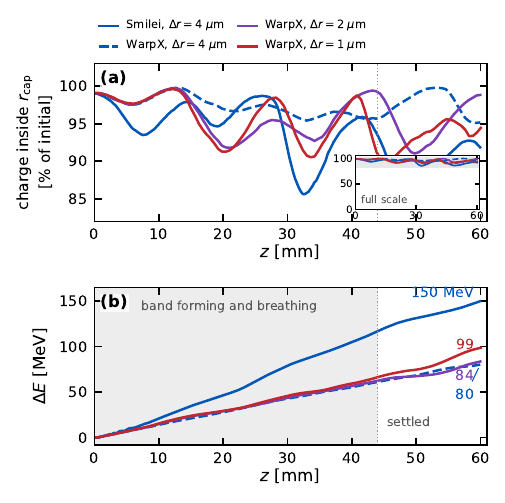}
\caption{\label{fig:transport}Transport of the captured witness ensemble in Smilei and WarpX.
The baseline Smilei calculation and the WarpX calculation on the same $\Delta r=4\;\mu$m grid are shown by solid and dashed blue curves; purple and red curves show WarpX at $\Delta r=2$ and $1\;\mu$m.
(a) Charge inside the capture radius, relative to the injected charge; the inset shows the full vertical scale.
(b) Captured-ensemble energy gain.
The gray shading covers band formation and breathing, ending at the dotted line at $z=44$~mm when the settled era begins.
The runs maintain high capture but differ in accumulated energy gain, with most of the baseline-to-WarpX difference established before settling.}
\end{figure}

The return geometry and field overlap establish a candidate accelerating structure. To determine whether this structure can successfully transport a bunch through the evolving wake, we track positron witnesses. The baseline simulation carries a low-charge witness through $60$~mm, resulting in $91.9\%$ capture and a $150$~MeV energy gain. Cross-code and resolution tests support the conclusion of high capture and phase-dependent transport, but they do not yield a converged energy gain (Table~\ref{tab:convergence}). To interpret these results, we relate the observed transport to the resolved force profile and quantify the associated emittance growth and energy spread. The injection history is critical throughout this process: the witness is initialized at $t=0$, before the wake forms, at the phase subsequently occupied by the settled band (Sec.~\ref{sec:law:bridge}), rather than being injected into a pre-existing wake. Detailed run configurations are listed in Appendix~\ref{app:impl:runs}, and beam loading effects are treated separately in Sec.~\ref{sec:witness:loading}.

\subsection{A stage of transport}
\label{sec:witness:baseline}

The baseline witness has $\gamma_w=5000$, a total charge of $0.1$~pC, and an initial placement of $\zeta_w=-227\;\mu$m. After $200$~ps (corresponding to $60$~mm of propagation), $91.9\%$ of its initial charge resides inside the capture radius $\rcap$, and the captured ensemble has gained $\Delta\gamma=+293$, or $150$~MeV [Fig.~\ref{fig:transport}(a,b)]. The stage-averaged gradient is $2.50$~GeV/m. A linear fit over the final $30$~ps yields $1.9$~GeV/m, indicating that acceleration continues at the simulation endpoint. The practical stage length of $50$ to $60$~mm is determined by the declining driver: the near-driver wake amplitude $A_{\rm conv}$ falls below $0.10\,E_0$ near $50$~mm. However, crossing this threshold does not mark the end of witness acceleration. The scaling with respect to driver energy is discussed in Sec.~\ref{sec:scaling}.

The captured fraction fluctuates with the transverse envelope, ranging from $86\%$ to $99\%$ after the first $5$~ps and stabilizing between $86\%$ and $96\%$ over the final $80$~ps. Longitudinally, the charge-weighted centroid of the full witness slips by $-0.19\;\mu$m, which is within the $0.23\;\mu$m kinematic bound. Each of the $2\times10^4$ tracked particles slips backward. The usable band evolves around this nearly stationary bunch. Its settled edges are $\zeta=-233.9\pm1.7\;\mu$m and $-212.1\pm1.0\;\mu$m (mean $\pm$ standard deviation over fourteen snapshots at $z=44$ to $60$~mm). This aligns closely with the final central $95\%$ captured-charge footprint, which spans $[-234.4,-219.4]\;\mu$m.

For intermediate comparisons at $30$~mm ($99$~ps), the baseline values are $93\%$ capture, $\Delta\gamma=+153$, and $\varepsilon_n=65$~mm\,mrad, which differ from the full-stage values of $91.9\%$, $+293$, and $83$~mm\,mrad.

\subsection{Resolution dependence and cross-code comparison}
\label{sec:witness:convergence}

We repeated the $60$~mm stage in WarpX~\cite{Vay2018}, release 26.07~\cite{WarpX2607}, using cylindrical geometry and radial cell sizes of $4$, $2$, and $1\;\mu$m.
The coarsest run matches the baseline grid, while the finer runs use $\Delta z=1\;\mu$m.
The same physical parameters and witness diagnostics allow the code and resolution dependences to be examined separately (Table~\ref{tab:convergence}).

\begin{table}
\caption{\label{tab:convergence}The same $60$~mm stage simulated in two codes at three radial resolutions, including a grid-matched pair at $\Delta r=4\;\mu$m. Capture is relative to the initial charge; energy and emittance are charge-weighted over the captured ensemble at the final time-series output near $t=200$~ps. The contrast row specifies the difference in capture percentage between the main witness and a control bunch placed $18\;\mu$m behind it, demonstrating position-sensitive confinement.}
\begin{ruledtabular}
\begin{tabular}{lcccc}
 & quasi-3D & \multicolumn{3}{c}{independent code} \\
 & $\Delta r=4$ & $\Delta r=4$ & $\Delta r=2$ & $\Delta r=1\;\mu$m \\
\hline
capture & $91.9\%$ & $95.2\%$ & $98.8\%$ & $94.5\%$ \\
$\Delta\gamma$ & $+293$ & $+157.2$ & $+163.7$ & $+193.2$ \\
$\Delta E$ [MeV] & $150$ & $80$ & $84$ & $99$ \\
$\varepsilon_n$ [mm\,mrad] & $83.3$ & $68.2$ & $76.0$ & $82.7$ \\
contrast [pct. points] & $55$ & $29.2$ & $44.0$ & $46.5$ \\
\end{tabular}
\end{ruledtabular}
\end{table}

Capture remains high across the four runs ($91.9\%$--$98.8\%$), with a $29$--$55$ percentage-point advantage over the rearward control bunch.
Emittance is more resolution dependent: it rises across the WarpX refinement sequence, with the finest result near $83$~mm\,mrad, similar to the baseline.
The finer force structure therefore matters for beam quality even where capture changes little.

Energy gain remains code dependent: the baseline yields $150$~MeV, compared with $80$, $84$, and $99$~MeV on successively finer WarpX grids.
The factor of $1.9$ difference at matched resolution shows that cell size alone does not explain the discrepancy.
The increasing refinement increments also prevent a continuum extrapolation from these three grids.

The gain histories separate two effects.
About two-thirds of the final baseline--fine-grid difference has accumulated by $30$~mm, whereas the WarpX refinement dependence develops mainly late in the stage.
The integrated cross-code gap thus largely reflects early wake evolution, even though the finest WarpX grid accelerates faster during the settled interval (Appendix~\ref{app:sens:branch}).

\subsection{The full stage in three dimensions}
\label{sec:witness:full3d}

A full Cartesian 3D run tests the $60$~mm stage without azimuthal truncation, retaining the baseline transverse cell size.

The working band lies further behind the driver in 3D, so the witness must follow its phase.
A control bunch at the baseline coordinate $\zeta=-227\;\mu$m retains $98.8\%$ of its charge but decelerates by $\Delta\gamma=-195$ over $30$~mm.
Placing the witness at $-250\;\mu$m instead preserves approximately the same distance behind the band front and yields accelerating transport.

The repositioned witness gains $\Delta\gamma=389$ ($199$~MeV) over $59.7$~mm, with final capture of $85.4\%$ and emittance of $60.6$~mm\,mrad.
Beyond $10$~mm, capture stays between $83.9\%$ and $88.5\%$, and the centroid remains within $0.5\;\mu$m.
The band settles near $39$~mm; its final edges are $24$--$27\;\mu$m behind the baseline edges, consistent with the early displacement of the band front.
The 3D result therefore demonstrates sustained acceleration at the shifted phase.
Its different placement and coarse grid leave the cross-code energy-gain discrepancy unresolved.

\subsection{Resolved focusing structure and limits of the field criterion}
\label{sec:witness:fine}
\label{sec:witness:aperture}

\begin{figure}
\includegraphics[width=\columnwidth]{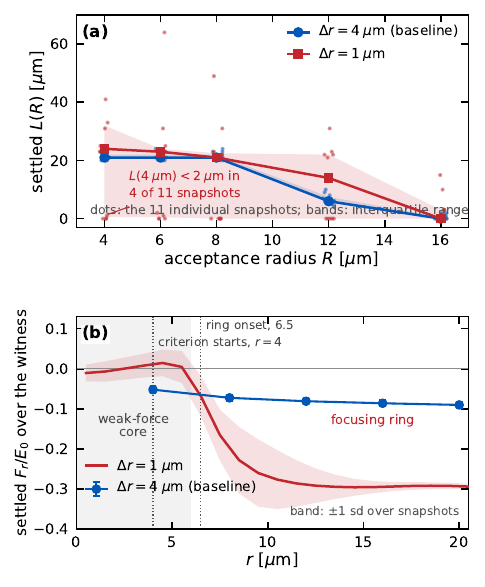}
\caption{\label{fig:aperture}The settled era in the baseline Smilei run ($\Delta r=4\;\mu$m) and the fine-grid WarpX run ($\Delta r=1\;\mu$m). (a) Usable length $L(R)$ from eleven snapshots per run at $160$ to $200$~ps, evaluated with the same criterion. Dots denote individual snapshots; lines and shading denote medians and interquartile ranges. The fine-grid median does not represent a persistent band. (b) Radial wake force on a positron averaged over the witness charge footprint, and subsequently over twelve fine-grid and thirteen baseline settled snapshots. Shading and error bars denote $\pm1$ standard deviation. The fine grid resolves a weak-force core inside $6\;\mu$m and a focusing ring beyond $6.5\;\mu$m. The criterion's initial force-sign test at $r=4\;\mu$m falls inside the core, where the two resolutions disagree.}
\end{figure}

The $1\;\mu$m field-diagnostic run resolves a weak-force core surrounded by a focusing ring [Fig.~\ref{fig:aperture}(b)].
Across the witness footprint, the mean force satisfies $|F_r|\lesssim0.014\,E_0$ inside roughly $6\;\mu$m, then becomes increasingly focusing from $-0.07\,E_0$ at $6.5\;\mu$m to $-0.29\,E_0$ at $12.5\;\mu$m.
At the criterion's starting radius of $4\;\mu$m, the baseline instead gives $F_r\simeq-0.05\,E_0$.
The local force-sign test consequently changes with resolution inside the core, while the surrounding ring remains clearly focusing.

This structure explains why the fine-grid usable interval can appear and disappear while transport persists.
The median lengths are $L(4)=24\;\mu$m and $L(6)=23\;\mu$m, but individual snapshots need not contain a continuous interval around the witness [Fig.~\ref{fig:aperture}(a)].
Over the same settled interval, capture rises from $91.8\%$ to $94.4\%$, and the mean Lorentz factor rises from $5141$ to $5189$.
The local field criterion is therefore a useful description of field overlap, but transport also depends on particle motion through the surrounding focusing ring.
The snapshot statistics are given in Appendix~\ref{app:sens:finegrid}.

The core radius is also smaller than the capture radius, $\rcap=21.5\;\mu$m.
In the fine-grid transport run, the instantaneous charge fraction inside $r<6\;\mu$m oscillates between $0.068$ and $0.538$, falling as the transverse envelope expands.
Particles can therefore move between the weak-force core and the focusing ring while remaining captured; core occupancy and captured charge describe different aspects of the bunch.

\subsection{Emittance}
\label{sec:witness:emittance}

\begin{figure}
\includegraphics[width=\columnwidth]{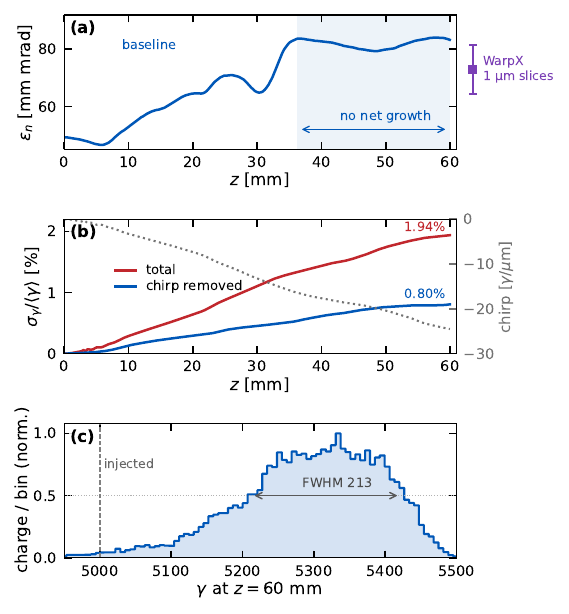}
\caption{\label{fig:quality}Beam quality of the baseline captured ensemble, with a WarpX slice comparison in (a).
(a) The normalized emittance grows most strongly between $30$ and $36$~mm and shows no net growth over the shaded final $24$~mm.
The separate purple point gives the range and median of six equal-charge slices at the endpoint of the $1\;\mu$m WarpX run.
(b) Relative rms energy spread before and after removal of the linear $\zeta$--$\gamma$ chirp.
The dotted curve (right axis) is the chirp slope; the residual reaches $0.80\%$ at the endpoint.
(c) End-of-stage spectrum of the baseline captured ensemble, with a peak at $\gamma=5334$ and a full width at half maximum of $213$.}
\end{figure}

In the baseline simulation, the captured-ensemble normalized emittance grows from nominally $50$ to $83$~mm\,mrad [Fig.~\ref{fig:quality}(a)].
The largest growth step occurs between $30$ and $36$~mm, while the band is still breathing.
Afterward, the emittance fluctuates near $83$~mm\,mrad with no net growth over the final $24$~mm.
The growth is therefore concentrated in the evolving wake, with a quieter interval after settling.

The $1\;\mu$m WarpX run also delivers a projected emittance near $83$~mm\,mrad.
Its six equal-charge longitudinal slices have transverse emittances of $64$ to $82$~mm\,mrad, with a median of $72.9$ [Fig.~\ref{fig:quality}(a)].
Much of the projected emittance is thus present within individual slices.
The nearly equal eigen-emittances and projected emittances indicate little inflation from transverse coupling (Appendix~\ref{app:impl:emittance}).

Reducing the injected emittance does not yield a proportionally lower delivered emittance.
On the same $1\;\mu$m WarpX grid, reducing $\varepsilon_n$ from $50$ to $10$~mm\,mrad and $\sigma_r$ from $7$ to $3.13\;\mu$m yields a final emittance of $38.0$~mm\,mrad: a factor of $2.18$ improvement for a fivefold reduction at injection.
Capture improves from $94.45\%$ to $99.99\%$, and the captured energy gain increases from $\Delta\gamma=+193.2$ to $+217.7$.
The smaller input beam therefore improves capture and delivery, but undergoes greater fractional emittance growth.

The low-emittance beam remains near $10$ to $12$~mm\,mrad over the first $10$~mm, then grows to approximately $29$~mm\,mrad near $20$~mm and $38.0$ at the endpoint.
This delayed growth again points to the evolving transport conditions as relevant to beam quality.
A $0.5\;\mu$m radial-grid control follows a similar evolution through $26.3$~mm, with emittances differing by about $15\%$; its shorter duration leaves the final emittance's resolution dependence unresolved (Appendix~\ref{app:impl:emittance}).

\subsection{Energy spread and the correlated chirp}
\label{sec:witness:spread}

The baseline captured ensemble exhibits a relative rms energy spread of $0.32\%$ at $36$~ps, $1.09\%$ at $99$~ps, and $1.94\%$ at $200$~ps [Fig.~\ref{fig:quality}(b)]. At the endpoint, the mean energy is $\langle\gamma\rangle=5293.2$ with an absolute spread of $\sigma_\gamma=102.6$. A linear $\zeta$--$\gamma$ chirp of $-24.5\;\gamma/\mu$m over the bunch length $\sigma_\zeta=3.81\;\mu$m contributes an rms spread of $93$. Removing this chirp leaves an uncorrelated regression residual of $42.6$, or $0.80\%$. This decomposition identifies a potentially correctable correlation in Fig.~\ref{fig:quality}(b), although downstream chirp compensation is not explicitly simulated here. The energy spectrum peaks at $\gamma=5334$ with a full width at half maximum (FWHM) of $213$, representing $4.0\%$ of the peak [Fig.~\ref{fig:quality}(c)].

Re-selecting the ensemble has little effect on the baseline total spread: a fixed-identity ensemble yields a nearly identical value of $2.01\%$. The code comparison shows a more substantial difference; the finest WarpX transport run yields a $3.2\%$ total spread and a steeper chirp of $-38.9\;\gamma/\mu$m, corresponding to a slice-median spread of $0.92\%$. That slice statistic and the baseline's $0.80\%$ regression residual are related diagnostics, but they are not identical estimators and should not be interpreted as a strict cross-code convergence test.

\section{Azimuthal stability, tolerances, and the solenoid}
\label{sec:stability}

The preceding transport results rely on a prescribed magnetic field and controlled beam placement. To evaluate departures from these ideal conditions, we test seeded driver asymmetries and witness placement errors before addressing the solenoid requirements. The seeded wake survives over the simulated interval, demonstrating finite-length robustness, although this does not establish long-distance stability or magnetic suppression of hosing. For the solenoid, analytical estimates and a targeted entrance-state test constrain specific effects, while an integrated magnet design and longitudinal-ripple tolerances remain tasks for future work.

\begin{figure}
\includegraphics[width=\columnwidth]{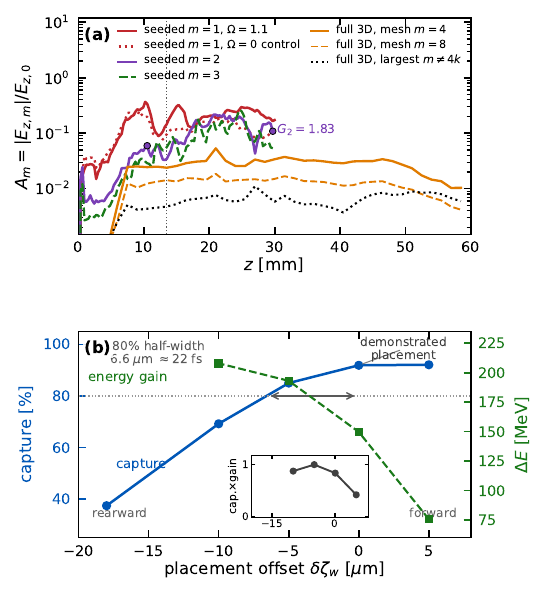}
\caption{\label{fig:stability}Wake response to driver asymmetry and witness placement.
(a) Normalized azimuthal amplitudes $A_m=|E_{z,m}|/E_{z,0}$ versus propagation distance.
Red curves show the response to a $2\;\mu$m driver dipole offset at $\Om=1.1$ (solid) and $\Om=0$ (dotted).
Purple and green curves show the response to $2\%$ quadrupole and sextupole seeds; the two circles on the quadrupole curve mark the $35$ and $99$~ps samples used for its growth factor $G_2$.
Orange curves show the $m=4$ and $8$ mesh harmonics in the unseeded full-3D run, and the black dotted curve shows its largest remaining mode, the dipole.
The dipole and full-3D curves sample the working bucket; the quadrupole and sextupole curves use the broader wake window specified in Appendix~\ref{app:sens:stability}.
The vertical dotted line marks the end of the early interval before appreciable driver-seed growth.
(b) Capture (left axis, five full-stage runs) and energy gain (right axis, four runs) versus longitudinal witness offset.
The rear-side half-width at $80\%$ capture is $6.6\;\mu$m ($22$~fs); the front-side limit is not bracketed.
The inset shows the capture--gain product normalized to its largest sampled value, at $\delta\zeta_w=-5\;\mu$m.}
\end{figure}

\subsection{Azimuthal content of the wake}
\label{sec:stability:modes}

The normalized azimuthal amplitudes $A_m\equiv|E_{z,m}|/E_{z,0}$ quantify departures from axisymmetry.
Seeded runs test the wake response to driver asymmetry over $30$~mm, while an unseeded full-3D run follows the stage through $60$~mm (Fig.~\ref{fig:stability}).
The analysis windows and initialization are specified in Appendix~\ref{app:sens:stability}.

A coherent $2\;\mu$m driver offset produces a dipole amplitude that peaks at $0.37$ near $34$~ps and falls to $0.17$ by the endpoint.
The unmagnetized control peaks at $0.25$ and ends at $0.10$.
Both wakes survive over $30$~mm, approximately two driver betatron periods ($\lambda_\beta=14.9$~mm), with a larger dipole response in the magnetized run.

With $2\%$ quadrupole and sextupole seeds, the corresponding wake amplitudes increase by factors of $1.83$ and $1.86$ between $35$ and $99$~ps, ending at $0.109$ and $0.053$.
Their transient maxima are larger: $A_2=0.22$ near $74$~ps and $A_3=0.26$ near $82$~ps.
The usable band survives these excursions even as the driver perturbations grow.

In the unseeded full-3D stage, the largest harmonics follow the Cartesian mesh symmetry: $A_4$ peaks at $0.053$ in the working bucket and $0.079$ in the deeper wake, while $A_8\le0.019$.
The remaining modes stay below $0.014$.
After peaking near $21$~mm, the working-bucket $A_4$ declines as the stage continues through settling.

These results establish finite-length wake robustness under the tested perturbations.
They do not show magnetic damping of driver asymmetry: the seeded driver dipole grows by factors of $6$--$10$ over $30$~mm, while the usable band survives.

\subsection{Placement and alignment}
\label{sec:stability:alignment}
\label{sec:witness:placement}

Because the band contracts over time, longitudinal witness placement represents a distinct constraint separate from azimuthal robustness. We quantify this using five $60$~mm runs that scan the offset from the baseline placement [Fig.~\ref{fig:stability}(b)]. At $\delta\zeta_w=+5$, $0$, $-5$, $-10$, and $-18\;\mu$m, the endpoint capture rates are $92.1\%$, $91.9\%$, $85.0\%$, $69.2\%$, and $37.4\%$, respectively. Interpolating the $80\%$ capture threshold yields a rear-side half-width of $6.6\;\mu$m, corresponding to $22$~fs at $\gamma_w=5000$. The front-side limit is not bracketed by the scan, as capture remains at $92.1\%$ at $+5\;\mu$m.

Moving the witness rearward trades captured charge for energy gain.
The offsets $+5$, $0$, $-5$, and $-10\;\mu$m give $\Delta\gamma=148$, $293$, $378$, and $407$, respectively.
The capture--gain product is largest at the sampled placement $-5\;\mu$m.
Relative to the baseline, the $-5$ and $-10\;\mu$m placements increase per-particle gain by approximately $29\%$ and $39\%$, while reducing capture to $85\%$ and $69\%$.

Transverse witness offsets of $3.5$ and $7\;\mu$m reduce capture by $0.4$ and $3$ percentage points, respectively.
A $1$~mrad tilt excites a $5\;\mu$m centroid oscillation while retaining $90.5\%$ capture and a final emittance of $82.4$~mm\,mrad, close to the untilted value.
The residual centroid offset of $1.29\;\mu$m indicates that a coherent component survives to the endpoint.

Driver-angle errors and solenoid tilt still require full 3D tests, as neither these witness-offset tests nor the seeded $m=1$ response at a fixed field geometry determine those specific tolerances. The complementary plasma-temperature tests detailed in Appendix~\ref{app:limits} address the forming structure at specified epochs, rather than full-stage witness capture or energy gain.

\subsection{The magnet}
\label{sec:stability:magnet}

The baseline uses a uniform axial field of $\Om=1.1$, corresponding to approximately $35$~T at $\n0=10^{16}\;\mathrm{cm^{-3}}$. Entrance state, uniform-field calibration, and hardware integration impose distinct constraints.

\subsubsection{Entrance and exit fringes}
The simulations initialize the plasma inside a uniform field and do not track it through a fringe. For a beam entering an axisymmetric solenoid from a field-free region with zero initial canonical angular momentum, Busch's theorem~\cite{Busch1926,Reiser2008} fixes the mechanical azimuthal momentum within the uniform field. Its magnitude is given by $|p_\phi|=eB_zr/2$, or $|p_\phi|/m_ec=\Om k_pr/2$, where the sign depends on the particle charge. This conservation constraint dictates the magnitude but does not determine profile-dependent aberrations or the full entrance-to-exit transfer map.

At the witness rms radius of $7\;\mu$m, the characteristic fringe kick is $0.072\,m_ec$, small compared with its intrinsic transverse momentum spread of $7.14\,m_ec$.
The corresponding driver kick at $35\;\mu$m is $0.36\,m_ec$.
Inside the ion column, the driver's solenoidal focusing is weak: $k_L^2/k_\beta^2\simeq6\times10^{-4}$.
Its entrance state matters more at the driver head, ahead of the ion column, where it can change the expulsion phase and hence the image position.

An entrance-state test gives the driver the azimuthal velocity acquired on entering the solenoid and follows it over $18.9$~mm.
The head contracts slightly relative to the $p_\phi=0$ reference, and the first image moves backward by approximately $7$--$10\;\mu$m.
The entrance condition can thus shift the image through its effect on the driver head.
The matched-time comparison and vacuum estimate are given in Appendix~\ref{app:sens:magnet}.

\subsubsection{Uniform-field calibration}
\label{sec:law:ripple}

At fixed anchor and calibrated wavelength, the positioning law gives $\mathrm{d}\zeta_1/\mathrm{d}\Om=\lambda_{p,\mathrm{eff}}/\Om^2$.
At $\Om=1.1$, a uniform field error therefore shifts the early image by $\delta\zeta_1\simeq328\;\mu\mathrm{m}\times(\delta B/B)$.
Keeping that shift within $5\;\mu$m requires approximately $1.5\%$ field calibration.
This estimate concerns the early density image; full-stage capture depends on the evolving working band.
Longitudinal field ripple requires integrating the gyro-phase along the electron orbit.

\subsubsection{Field strength and integration}

The simulated wake occupies radii below approximately $50\;\mu$m over a $60$~mm stage.
A magnet must additionally accommodate the beamline, plasma source, and diagnostics.
Laboratory magnets combining superconducting coils with resistive background fields have demonstrated fields above $35$~T~\cite{Hahn2019}; integrating that field strength with a plasma stage requires a suitable bore, field volume, and fringe geometry.
Density scaling trades these requirements against larger stage dimensions (Sec.~\ref{sec:scaling}).

\section{Beam loading and extraction efficiency}
\label{sec:witness:loading}

Increasing the witness charge tests how much energy can be extracted from the magnetically organized wake and how the beam quality changes in the process.
We follow this response from weak loading through overload, then examine how current-profile shaping interacts with the finite usable band.

\subsection{Low-charge response}
\label{sec:loading:lowcharge}

Starting from the $0.1$~pC baseline, whose peak witness density is $0.02\,\n0$, we conduct repeat simulations at $0.5$, $2.5$, and $10$~pC, with the latter reaching a peak density of $2.02\,\n0$.
Despite this substantial density increase, the median usable length $L(4\;\mu\mathrm{m})$ remains $46$, $47$, and $48\;\mu$m, respectively (compared to $46\;\mu$m for the $0.1$~pC witness).
The in-band peak accelerating field at $r=4\;\mu$m also remains consistently at $0.663\,E_0$ to three significant digits.
At $99$~ps ($30$~mm), capture rates are $93.5\%$, $93.5\%$, and $93.6\%$, with emittances differing by only $0.3\%$.
The $10$~pC captured-ensemble gain drops by $\Delta\gamma=4.0$ relative to the $0.5$~pC value, representing an approximately $3\%$ reduction from the initial $+153$.

The limited wake perturbation observed here is consistent with the short response time available during the bunch passage. The at-bunch density perturbations are $0.0027$, $0.0160$, and $0.0665\,\n0$, agreeing with linear response theory to within $15\%$. The small scale parameter $(k_p\sigma_z)^2=0.0057$ limits the plasma-electron response while the bunch passes, and these runs do not exhibit nonlinear core filling. However, this short-bunch result need not persist at higher densities if the physical bunch length remains fixed.

The $0.1$~pC witness gains only about $0.02\%$ of the driver energy loss over the full $60$~mm stage.
Increasing the witness charge is therefore essential for energy extraction, motivating the broader loading scan below.

\subsection{Charge dependence and extraction efficiency}
\label{sec:loading:efficiency}

\begin{figure}
\includegraphics[width=\columnwidth]{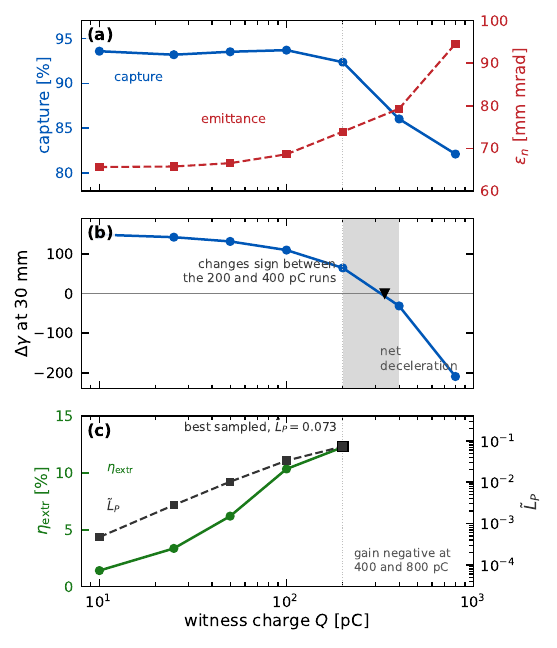}
\caption{\label{fig:loading}The extended loading scan, evaluated at the common $99$~ps ($30$~mm) epoch. (a) Capture and normalized emittance of the captured ensemble at $30$~mm. (b) Energy gain: the shaded bracket indicates the region where the gain changes sign (between the $200$ and $400$~pC runs), and the triangle represents the linear interpolation between them. (c) Extraction efficiency (left axis) and the dimensionless luminosity per unit power (right axis, logarithmic scale) derived from Eq.~(\ref{eq:lumi}), calculated for charges that yield a positive gain. The large square designates the best sampled point at $200$~pC. Dotted vertical lines mark the best sampled charge, $200$~pC.}
\end{figure}

Increasing witness charge raises extraction while reducing the energy gained per captured particle [Fig.~\ref{fig:loading}].
At $99$~ps ($30$~mm), capture remains $92\%$--$94\%$ through $200$~pC, while emittance rises from $65.6$~mm\,mrad at $10$~pC to $73.9$~mm\,mrad at $200$~pC.
The captured gain falls from $\Delta\gamma=142$ at $25$~pC to $65$ at $200$~pC and changes sign between $200$ and $400$~pC.
At $800$~pC, the overloaded witness decelerates by $\Delta\gamma=-209$, capture falls to $82\%$, and emittance rises to $95$~mm\,mrad.

We characterize extraction and beam quality using the efficiency $\eta_{\rm extr}$ and the dimensionless luminosity per unit power $\tilde L_P$ of Cao et al.~\cite{Cao2024}, Eq.~(11), evaluated from these runs [Fig.~\ref{fig:loading}(c)]:
\begin{equation}
\tilde L_P=4\pi r_e\,\frac{\eta_{\rm extr}N}
{\sqrt{\varepsilon_{nx}\varepsilon_{ny}}},
\label{eq:lumi}
\end{equation}
where $r_e$ is the classical electron radius and $N=Q/e$ is the injected witness population. The efficiency rises from $1.4\%$ at $10$~pC to $12.3\%$ at $200$~pC, while $\tilde L_P$ increases from $4.7\times10^{-4}$ to a maximum sampled value of $0.073$ at $200$~pC. This value is approximately one order of magnitude below the $0.4$ to $0.9$ range cited for the best published positron schemes.

Here, $\eta_{\rm extr}$ includes the energy change of the entire witness bunch, while $\tilde L_P$ combines the injected population with the captured-ensemble emittance.
At $30$~mm, the best sampled value is $\tilde L_P\simeq0.07$; extraction and beam-quality diagnostics are detailed in Appendix~\ref{app:sens:loading}.

At $200$~pC, $\eta_{\rm extr}/(k_p\sigma_z)=1.63$ is consistent with the order-unity short-bunch loading scale discussed in Ref.~\cite{Cao2024}.
The useful-charge limit in this scan appears through the loss of accelerating field before capture collapses.
The diagnostic checks and comparison with electron-motion and magnetization thresholds are given in Appendix~\ref{app:sens:loading}.

\subsection{Longitudinal field and transverse response}
\label{sec:loading:fields}

For the Gaussian witness, increasing charge mainly lowers the accelerating field without flattening its longitudinal profile.
In a fixed window through the settled band, the mean field falls from $0.410\,E_0$ at $0.1$~pC to $0.386$, $0.361$, $0.310$, and $0.206\,E_0$ at $25$, $50$, $100$, and $200$~pC.
It reaches $0.007\,E_0$ at $400$~pC and $-0.321\,E_0$ at $800$~pC, with an approximately linear response through $400$~pC.
The absolute peak-to-peak ripple is $0.096\,E_0$ at $0.1$~pC and $0.103\,E_0$ at $200$~pC.
As the mean field decreases, this nearly unchanged ripple grows from $23\%$ to $50\%$ of the mean, coupling increased extraction to worsening relative field uniformity.
The averaging convention is specified in Appendix~\ref{app:sens:loading}.

The transverse response changes less than the accelerating field over the moderate-charge scan.
The baseline-grid core spring estimator varies little through $400$~pC and increases at $800$~pC (Appendix~\ref{app:sens:loading}).
Substantial longitudinal loading therefore precedes a large change in the core spring estimate; the fine-grid force profile is discussed in Sec.~\ref{sec:witness:fine}.

\subsection{Witness-profile shaping}
\label{sec:loading:shaping}

A $170$~pC shaped witness tests a longitudinal profile designed from the simulated linear wake response (Appendix~\ref{app:sens:kernel}).
Compared with a $200$~pC Gaussian witness, this configuration gives a $41\%$ higher design-profile-weighted mean field, with a relative field spread of $12.95\%$ versus $12.73\%$.
The field diagnostic weights the time-averaged on-axis field by the design profile.

At the $60$~mm endpoint, the shaped bunch has $84.3\%$ capture, compared with $90.8\%$ for the Gaussian bunch.
Its mean gain is larger ($\Delta\gamma=161.8$ versus $125.6$), but its rms spread relative to that gain, $\sigma_\gamma/\langle\Delta\gamma\rangle$, is larger by a factor of $2.4$.
The product $Q\langle\Delta\gamma\rangle\,\mathrm{capture}$ differs by only a factor of $1.015$.

The increased mean field does not produce improved flatness or delivered charge.
The shaped bunch carries less charge overall, and its injected profile places $30\%$ of the charge behind the mean rear edge of the settled band, compared with $4\%$ for the Gaussian.
This result makes the longitudinal support of the profile a design constraint alongside its field response.
Current-profile shaping must account jointly for beam placement, charge distribution, and the accelerating and confining region encountered during transport.

\section{Scaling with driver energy and plasma density}
\label{sec:scaling}

Raising the driver energy can extend the acceleration distance without requiring a stronger magnetic field. Conversely, raising the plasma density increases both the accelerating field scale $E_0$ and the required magnetic field, scaling as $B \propto \sqrt{\n0}$ at a fixed magnetization parameter $\Om$. We first compare the baseline configuration with a fivefold increase in driver energy, then discuss conditional density scaling, and conclude by examining the limits of multistage extrapolation and driver-species transfer.

\subsection{Driver energy}
\label{sec:scaling:energy}

\begin{figure*}
\includegraphics[width=\textwidth]{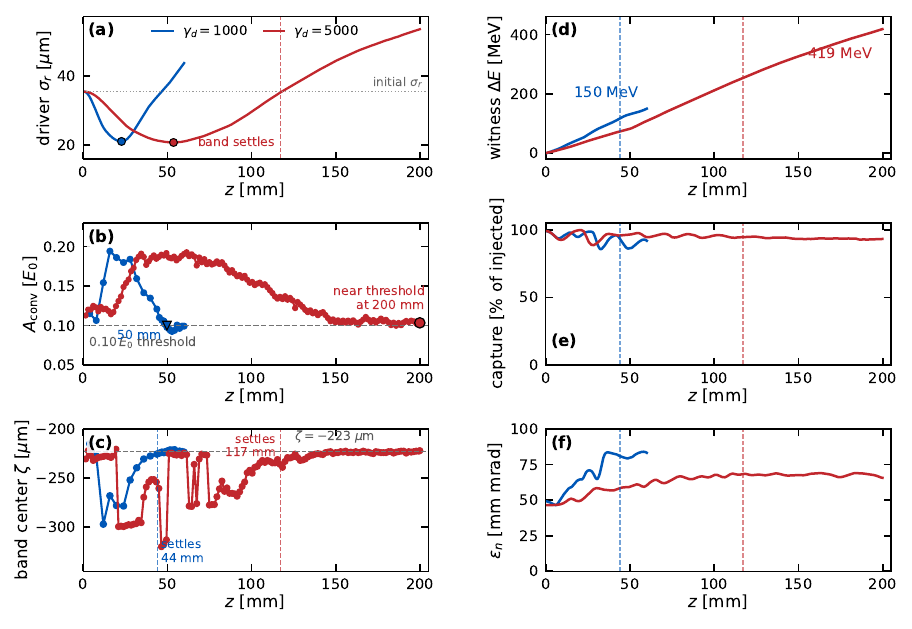}
\caption{\label{fig:gscaling}Driver-energy comparison versus propagation distance, showing the baseline $\gamma_d=1000$ run and the continuous $\gamma_d=5000$ run extended to $200$~mm while carrying the main and rearward-control witnesses. The left column displays: (a) driver rms radius, with circles marking its minimum in each run, (b) near-driver amplitude $A_{\rm conv}$ based on a common estimator (Appendix~\ref{app:impl:amp}), and (c) band center. The triangle in (b) marks the baseline's amplitude-threshold crossing, which serves as an operational metric rather than the absolute end of acceleration. The right column displays the main-witness parameters: (d) energy gain, (e) capture relative to the injected charge, and (f) normalized emittance. Energy and emittance are charge-weighted over the instantaneous captured ensemble. Colored dashed vertical lines mark the band settling points at $44$ and $117$~mm. Each curve terminates at its respective run's endpoint; the higher-energy witness continues gaining energy after settling without experiencing further net emittance growth.}
\end{figure*}

To isolate driver-energy effects from changes in magnetization, the higher-energy calculation retains the baseline plasma density, magnetic field, and initial driver profile, increasing only $\gamma_d$ from $1000$ to $5000$. A driver-only calculation extends to $150$~mm, while a separate continuous run carrying the witness pair reaches $200$~mm [Fig.~\ref{fig:gscaling}]. Field diagnostics characterize the wake evolution, and tracked particles determine the resulting gain and emittance.

The higher-energy driver contracts from $35.4\;\mu$m to $20.8\;\mu$m rms at $53.6$~mm, compared with the baseline pinch at $22.8$~mm.
The pinch distance increases by a factor of $2.4$, close to the $\sqrt{\gamma_d}$ betatron scaling ($\sqrt5\simeq2.24$).
Band settling occurs later, at approximately $117$~mm rather than $44$~mm, a factor of $2.7$.
Increasing the driver energy therefore stretches the formation transient, with distinct distances for maximum compression and for the settling of the field overlap.

Over $z\ge150$~mm, the higher-energy band has a center of $-223.4\pm0.7\;\mu$m and a length $L(4\;\mu\mathrm{m})=21.2\pm1.6\;\mu$m (mean $\pm$ temporal standard deviation, $34$ snapshots).
These agree with the baseline center of $-223.0\pm1.2\;\mu$m and median length of $21\;\mu$m.
The two energies thus produce similar late-time field overlap despite different formation distances.
The early-image positioning law accounts for the magnetic return scale; predicting the settled phase additionally requires the evolving driver and wake fields.

The higher-energy wake strengthens during driver compression: the near-driver amplitude $A_{\rm conv}$ reaches $0.193\,E_0$ at $61$~mm, shortly after the pinch, and then declines.
The driver re-expands through its initial radius near $118$~mm, close to band settling, and continues expanding to $53.4\;\mu$m rms by $200$~mm while the band remains near $\zeta=-223\;\mu$m.
A stationary band therefore coexists with continuing changes in the driver size and wake amplitude.

The common amplitude diagnostic further separates band settling from the end of acceleration.
In the baseline, $A_{\rm conv}$ crosses $0.10\,E_0$ near $50$~mm and ends at $0.099\,E_0$ while the witness continues gaining energy.
In the higher-energy run it remains between $0.100$ and $0.111\,E_0$ over $150$--$200$~mm.
The demonstrated settled acceleration interval consequently extends for at least $83$~mm, from $117$~mm to the endpoint; the ultimate depletion distance lies beyond what these data determine.

The witness launched at $\zeta_w=-224\;\mu$m retains $93.2\%$ capture at $200$~mm and gains $\Delta\gamma=819$ ($419$~MeV).
Its mean gradient is $1.97$~GV/m over the settled interval from $117$ to $200$~mm, declining to $1.8$~GV/m over the final $24$~mm.
The captured-ensemble emittance rises from $46.5$ to $65.7$~mm\,mrad over the full stage, but decreases slightly from $68.4$ to $65.7$~mm\,mrad after settling.
Increasing the driver energy therefore extends sustained acceleration without further net emittance growth during the settled phase.

\subsection{Plasma density and the magnet ceiling}
\label{sec:scaling:density}

At a fixed magnetization parameter $\Om=\wc/\wp$, the required magnetic field scales as
\begin{equation}
B=32\,\Om\sqrt{\n0/10^{16}\;\mathrm{cm^{-3}}}\;\;\mathrm{T}.
\label{eq:bscale}
\end{equation}
At $\Om=1.1$, plasma densities of $10^{16}$, $10^{17}$, and $10^{18}\;\mathrm{cm^{-3}}$ require approximately $35$, $112$, and $350$~T, respectively.
Under plasma similarity scaling, normalized beam and plasma parameters remain fixed, fields scale as $\sqrt{\n0}$, and lengths scale as $k_p^{-1}$.
The stronger gradient and shorter stage then compensate in their product, preserving the energy-gain scale.

For example, the baseline's approximately $21\;\mu$m settled band would contract to about $2\;\mu$m at $10^{18}\;\mathrm{cm^{-3}}$.
Scaling the witness length with it preserves $k_p\sigma_z$ and the short-bunch loading regime.
Keeping $\sigma_z=4\;\mu$m instead gives $k_p\sigma_z=0.75$ and a bunch longer than the scaled band, breaking similarity with the demonstrated configuration.
Reducing density relaxes the magnetic-field requirement while increasing the physical sizes of the stage, plasma, and bunch.

The plasma must also extend far enough transversely to support electron return.
The low-density column tests weaken the wake at a flat radius of $1.19\,c/\wp$ and lose its oscillatory structure at $0.37\,c/\wp$ (Appendix~\ref{app:limits:width}).
These geometries favor a broad plasma over a narrow filament; preserving their normalized width at lower density requires a wider physical plasma source.

\subsection{Staging}
\label{sec:scaling:staging}

Staging requires control of the formation transient as well as sustained acceleration within each stage.
Repeating a $50$ to $83$~mm\,mrad emittance increase multiplicatively would produce more than a hundredfold increase over ten stages.
That extrapolation makes interstage matching a central design question: the bunch entering a later stage need not repeat the first stage's mismatch and transient growth.
The present single-stage results provide the input beam and field scales for that study, with deliverable charge further constrained by the short usable band and beam loading.

\subsection{Driver species}
\label{sec:scaling:driver}

After expulsion, the ideal gyro clock depends on the magnetic field and plasma-electron dynamics, while the driver's charge sign affects how the electrons are launched.
The positron-driver calibration across seven field strengths yields a $1/\Om$ fit with rms residual $2.2\;\mu$m, and its fitted wavelength transfers to the electron-driver analysis (Sec.~\ref{sec:law:calibration}).
This agreement supports a common return-time scale, while the anchor, field overlap, and transport remain properties of the particular driver and wake.

\section{Comparison with other positron-acceleration mechanisms}
\label{sec:compare}

Positron-acceleration schemes differ in how they create overlapping accelerating and focusing fields.
The main choices are the wake regime, the plasma geometry, the driver profile, and the response induced by the witness.

\subsection{Quasi-linear wakes and hollow channels}

Quasi-linear wakes provide a positron-accelerating and focusing phase through modest plasma-electron displacements.
Doche et al.~\cite{Doche2017} demonstrated trailing-positron acceleration across nonlinear and quasi-linear regimes using a positron driver, with more regular wakefields in the quasi-linear case.
The magnetized scheme instead retains a nonlinear electron driver and reorganizes the returning sheath.

An ideal hollow plasma channel removes plasma from the beam path, giving a force-free interior for an on-axis relativistic witness and requiring external focusing~\cite{Gessner2016}.
Beam offsets excite transverse wakefields, as observed experimentally by Lindstr\"om et al.~\cite{Lindstrom2018}.
The channel therefore changes both the source of confinement and the response to misalignment.

\subsection{Shaping the driver, plasma, or witness}

A hollow, or donut-shaped, electron driver changes the transverse forcing of a uniform plasma.
The annular driver produces a wake with a region of simultaneous positron acceleration and focusing~\cite{Jain2015}.
Here the driver profile supplies the structural control; the magnetized configuration uses an axial field with a Gaussian electron driver.
Orbital-angular-momentum laser drivers provide another route to reshaping the sheath~\cite{Vieira2014}.

Finite-radius plasma columns modify the restoring force on expelled electrons.
Outside the ion column, the weaker restoring force spreads their return into an extended on-axis filament~\cite{Diederichs2019,Diederichs2020}.
Low-emittance transport and damping of misalignment oscillations have been demonstrated within this family of schemes~\cite{Diederichs2020,Diederichs2022}.
The plasma geometry establishes the focusing structure, while witness shaping and matching control beam quality.

An additional electron bunch can also extend electron accumulation into the positron-accelerating phase by loading the back of the primary bubble~\cite{Wang2021}.
Its charge and placement control the plasma response.
In self-loading schemes, the positron witness itself draws electrons into a focusing filament and modifies the accelerating field.
Zhou et al.~\cite{Zhou2025} studied this process in an electron-driven blowout, while Corde et al.~\cite{Corde2015} demonstrated a positron-driven self-loaded configuration.

\subsection{Magnetic control of the return}

The axial field controls the plasma-electron trajectories through the gyro clock and the CAM barrier (Sec.~\ref{sec:theory}).
Together they organize the longitudinal return phase and its transverse scale, forming an electron-rich structure in a uniform plasma before substantial witness loading.
The resulting plasma wakefields accelerate and confine the witness.

This separation between structure formation and loading makes the magnetic field an additional design parameter.
The loading scan then exposes the next constraint: extracting more energy lowers the accelerating field and changes beam quality, while profile shaping must fit the charge into the evolving usable band (Sec.~\ref{sec:witness:loading}).
The corresponding hardware requirement is a $35$~T baseline axial field, with $B\propto\sqrt{\n0}$ at fixed $\Om$; density scaling trades field strength against physical stage dimensions (Sec.~\ref{sec:scaling:density}).

\section{Conclusions}
\label{sec:conclusions}

Magnetizing the nonlinear electron-driven wakefield provides a robust route to positron acceleration. Rather than attempting to avoid the nonlinear blowout regime, an axial magnetic field reorganizes the returning plasma electrons into finite-radius gyro images. This structure is governed by two fundamental invariants: the quasi-static invariant regulates the longitudinal spacing (acting as an ideal gyro clock), while the conservation of canonical angular momentum restricts the radial collapse of the returning electrons toward the axis. The result is a quasi-achromatic, electron-rich column that provides overlapping regions of strong positron acceleration and transverse confinement.

Tracked particle simulations confirm that this magnetically organized wake successfully transports positrons. Operating at a $35$~T, $\n0=10^{16}\;\mathrm{cm^{-3}}$ baseline, a witness bunch injected at the settled phase achieves $92\%$ capture over a $60$~mm stage. Crucially, while the captured-ensemble emittance initially grows due to the wake formation transient, it saturates with no further net growth during the late stages of transport. Scaling to a higher driver energy ($\gamma_d=5000$) further demonstrates sustained transport, yielding $419$~MeV over an extended $200$~mm stage with $93\%$ capture and confirming the long-range persistence of the settled wake structure.

Despite these promising transport characteristics, key challenges remain. First, while independent particle-in-cell codes consistently verify sustained beam transport and high capture rates, the exact energy gain and final emittance exhibit strong sensitivity to numerical resolution. This highlights the demanding computational requirements for accurately modeling near-axis positron dynamics. Second, although the magnetized focusing structure forms independently of the witness bunch, it is ultimately susceptible to beam-induced wake degradation. High-charge extraction reduces the accelerating gradient and degrades beam quality---a fundamental trade-off that simple longitudinal profile shaping could not resolve in our tests.

Translating this physical mechanism into a practical collider stage will require evolving this single-stage demonstration into a fully matched, efficiently loaded accelerator. This entails precise control over the early formation transient, realistic modeling of fringe-field injection and extraction, and the engineering integration of high-field solenoids with the plasma source. Furthermore, simply increasing the plasma density to achieve higher gradients is strongly constrained by the steep magnetic scaling requirement ($B\propto\sqrt{\n0}$). Consequently, exploiting this magnetized blowout regime will require careful parameter optimization and interstage matching for future collider designs.

\begin{acknowledgments}
We thank the National Center for High-performance Computing (NCHC), Taiwan, for providing the Taiwania 3 supercomputer and computing resources.
PC appreciates the support by Taiwan's National Science and Technology Council (NSTC) under funding accounts: 115-2112-M-002-014- and 115-2221-E-002-189-MY3.
Work at SLAC is supported by the U.S. Department of Energy under contract DE-AC02-76SF00515.
\end{acknowledgments}

\section*{Data Availability}
\label{sec:methods:data}

The data supporting the findings of this study are available from the authors upon reasonable request.

\appendix
\section{Numerical implementation details}
\label{app:impl}

This appendix specifies the simulation setup, common field and witness diagnostics, and the inventory of transport runs.

\subsection{Grid, solver, and initialization}
\label{app:impl:solver}

\emph{Time step and mode truncation.} The Courant limit of the cylindrical solver becomes stricter as higher azimuthal modes are included. Production runs utilizing two modes ($m\le1$) operate at a time step of $0.9\,\Delta z/c$. Stability runs evaluating up to four modes ($m\le3$) require a reduced step of $0.75\,\Delta z/c$. The full-3D Cartesian simulations operate at $0.754\,\Delta x/c$.

\emph{External fields under a moving window.} In the quasi-3D simulations, the axial magnetic field is naturally applied as an external grid field in the $m=0$ component. However, in the full-3D Cartesian geometry, the uniform axial field is applied as a prescribed analytic field evaluated directly at the particle positions to ensure consistency under moving-window translation.

\phantomsection
\label{app:impl:quiet}
\emph{Quiet initialization.} The driver macroparticles use mirror-symmetric quiet-start sampling to reduce the unseeded odd azimuthal moments.
The deliberately applied centroid offset then supplies a controlled dipole seed for comparison with the quiet reference (Sec.~\ref{sec:stability:modes}).

\subsection{Diagnostic conventions}
\label{app:impl:diagnostics}

\emph{Cylindrical grids and staggering.} The first AM radial row is a ghost row and is excluded from physical sampling.
The density and transverse fields are stored at dual-grid coordinates, one half radial cell below their labels; $E_z$ uses the primal-grid labels.
The transverse fields are linearly interpolated to the primal radii before evaluating $F_r=E_r-B_\theta$ and the usability criterion.
The criterion begins at physical $r=4\;\mu$m across resolutions, and the quality scan tests focusing at physical $r=8\;\mu$m.

\phantomsection
\label{app:impl:amp}
\emph{Wake amplitude definitions.} We distinguish between two wake-amplitude diagnostics.
The near-driver amplitude ($A_{\rm conv}$) evaluates the maximum smoothed accelerating field strictly within the interval $-200\le\zeta\le-60\;\mu$m behind the driver, providing a metric to track the near-driver wake evolution (Sec.~\ref{sec:scaling:energy}).
Conversely, the bucket-global amplitude evaluates the absolute maximum $|E_z|$ across the entire first bucket, yielding a substantially higher value.

\subsection{Witness diagnostic conventions}
\label{app:impl:witness}

The baseline witness phase space is recorded every $1$~ps with per-particle weights; the high-resolution WarpX diagnostics operate at a $2$~ps cadence (Sec.~\ref{sec:witness:aperture}).
Witness statistics are charge-weighted over the \emph{captured ensemble}, defined dynamically at each output time as the particles residing within the capture radius $\rcap=21.5\;\mu$m.
Re-selection permits particles to leave and re-enter the ensemble.
To verify that this dynamic selection does not skew the metrics, a fixed-identity ensemble selected at $t=36$~ps ($r<10\;\mu$m) reproduces the re-selected moments to within a few parts per thousand.
The emittance and resolution comparisons are detailed in Appendix~\ref{app:sens:resolution}.

\subsection{Simulation inventory}
\label{app:impl:runs}

Table~\ref{tab:runs} provides a consolidated inventory of the witness-transport simulations.
\begin{table*}
\caption{\label{tab:runs}Inventory of witness-transport runs discussed in Sec.~\ref{sec:witness}. RZ denotes WarpX's cylindrical $r-z$ geometry; $\Delta r$ represents the transverse cell size in both cylindrical and Cartesian runs. ``As baseline'' indicates the standard witness parameters: $\gamma_w=5000$, $\varepsilon_n=50$~mm\,mrad, $\sigma_r=7\;\mu$m, and total charge $0.1$~pC.}
\begin{ruledtabular}
\begin{tabular}{lccp{0.25\textwidth}p{0.25\textwidth}}
Code and geometry & $\Delta r$ & Stage & Witness placement & Primary objective \\
\hline
Smilei, quasi-3D ($m\le1$) & $4\;\mu$m & $60$~mm & Main at $\zeta_w=-227\;\mu$m, rear control at $-245\;\mu$m & Baseline transport demonstration (Sec.~\ref{sec:witness}) \\
WarpX, RZ FDTD & $4\;\mu$m & $60$~mm & As baseline & Cross-code comparison at matched grid resolution \\
WarpX, RZ FDTD & $2\;\mu$m & $60$~mm & As baseline & Radial-resolution dependence \\
WarpX, RZ FDTD & $1\;\mu$m & $60$~mm & As baseline & Finest grid; mapping near-axis force structure (Sec.~\ref{sec:witness:fine}) \\
WarpX, RZ FDTD & $1\;\mu$m & $60$~mm & $\varepsilon_n=10$~mm\,mrad, $\sigma_r=3.13\;\mu$m & Beam-quality response to lower injected emittance \\
WarpX, RZ FDTD & $0.5\;\mu$m & $26.3$~mm & As above & Resolution control for the low-emittance run \\
Smilei, full-3D Cartesian & $4\;\mu$m & $30$~mm & Placed at absolute baseline $\zeta_w=-227\;\mu$m & Demonstrates absolute placement does not perfectly transfer to 3D \\
Smilei, full-3D Cartesian & $4\;\mu$m & $59.7$~mm & Placed at $\zeta_w=-250\;\mu$m (matched to band front) & Full-stage 3D transport without mode truncation \\
\end{tabular}
\end{ruledtabular}
\end{table*}

\section{Supporting analyses and sensitivity checks}
\label{app:sens}

This appendix collects the extraction methods, supporting diagnostics, and sensitivity checks for the wake structure, electron returns, witness transport, and stability.

\subsection{Band usability, field metrics, and evolution}
\label{app:sens:fields}

\emph{Usable length evolution.} Table~\ref{tab:snapshots} gives the band lengths and radial field nonuniformity at each scan-window snapshot, supporting the ranges quoted in Sec.~\ref{sec:wake:lofr}.
The usable length $L(R)$ varies with time.
Across these snapshots, the magnetized band is longer and its accelerating field is more uniform radially than in the unmagnetized case.

\begin{table}
\caption{\label{tab:snapshots}The aperture sweep and the island-averaged radial nonuniformity of $E_z$ between $r=4$ and $12\;\mu$m, at each of the four scan-window snapshots, with the field and without it.
Lengths in micrometers.
The latest snapshot is the one Fig.~\ref{fig:lofr} draws as curves and the one the abstract quotes.}
\begin{ruledtabular}
\begin{tabular}{lcccc}
$t$ [ps] & $43.1$ & $50.1$ & $57.1$ & $64.7$ \\
\hline
$L(4\;\mu\mathrm{m})$, $\Om=1.1$ & $59$ & $48$ & $54$ & $104$ \\
$L(4\;\mu\mathrm{m})$, $\Om=0$ & $25$ & $14$ & $22$ & $28$ \\
$L(6\;\mu\mathrm{m})$, $\Om=1.1$ & $59$ & $48$ & $47$ & $94$ \\
$L(6\;\mu\mathrm{m})$, $\Om=0$ & $21$ & $9$ & $19$ & $22$ \\
nonuniformity, $\Om=1.1$ & $9.1\%$ & $8.4\%$ & $8.7\%$ & $6.3\%$ \\
nonuniformity, $\Om=0$ & $40.7\%$ & $76.5\%$ & $36.1\%$ & $36.2\%$ \\
\end{tabular}
\end{ruledtabular}
\end{table}

\emph{The $\Om$ scan.} Table~\ref{tab:omscan} uses physical radial coordinates, with the transverse force interpolated onto the $E_z$ grid before evaluating the gates.
The quality-interval length peaks at $\Om=1.0$, while its width--field product peaks at $1.1$.
Requiring focusing at every sampled radius from $4$ to $16\;\mu$m gives maximum interval length at $\Om=1.0$ ($50\;\mu$m), compared with $41\;\mu$m at $1.1$.
Thus, the preferred field depends on both the aperture and the quantity being optimized.

\begin{table}
\caption{\label{tab:omscan}Field-strength scan with $11\;\mu$m longitudinal smoothing, scan-window time averaging and driver exclusion.
Columns 2--4 give the quality-interval length, its length with focusing also required at $r=4\;\mu$m, and its median accelerating field.
The last two columns give $E_z$ at $r=4\;\mu$m and the signed variation $[E_z(24)-E_z(4)]/E_z(4)$ at the band's working point; radii are in micrometers.  \emph{No band} denotes absence of simultaneous acceleration and focusing at $r=8\;\mu$m.
At $\Om=0.7$ and $1.7$, the band contains no interval satisfying the quality gates, so no quality-interval field is reported.}
\begin{ruledtabular}
\begin{tabular}{cccccc}
$\Om$ & quality & also $F_r<0$ & $E_z$ in it & $E_z(4)$ &
radial \\
 & interval [$\mu$m] & at $4\;\mu$m [$\mu$m] & [GV/m] & [$E_0$] & var. [\%] \\
\hline
$0.60$ & \multicolumn{5}{c}{no band} \\
$0.70$ & $0$ & $0$ & --- & $0.085$ & $+47.0$ \\
$0.80$ & $35$ & $6$ & $1.77$ & $0.132$ & $-14.2$ \\
$0.85$ & $59$ & $24$ & $2.20$ & $0.148$ & $-6.1$ \\
$0.90$ & $78$ & $43$ & $2.48$ & $0.161$ & $-0.7$ \\
$0.95$ & $89$ & $61$ & $2.72$ & $0.217$ & $+2.2$ \\
$1.00$ & $94$ & $84$ & $3.02$ & $0.281$ & $+3.4$ \\
$1.10$ & $87$ & $83$ & $3.47$ & $0.415$ & $+6.1$ \\
$1.20$ & $60$ & $60$ & $3.85$ & $0.424$ & $+12.6$ \\
$1.30$ & $36$ & $36$ & $4.06$ & $0.422$ & $+29.8$ \\
$1.40$ & $18$ & $18$ & $4.18$ & $0.416$ & $+47.9$ \\
$1.70$ & $0$ & $0$ & --- & $0.376$ & $+79.7$ \\
\end{tabular}
\end{ruledtabular}
\end{table}

\phantomsection
\label{app:sens:criterion}
\emph{Criterion sensitivity.} The robustness scan (Sec.~\ref{sec:methods:robustness}) evaluates the usability criterion across varying accelerating thresholds ($0.05$ to $0.15\,E_0$), uniformity tolerances, and smoothing kernels.
The magnetized-to-unmagnetized length contrast remains at least a factor of two across all 36 evaluated parameter combinations.
However, raising the baseline field threshold above $0.3\,E_0$ shifts the comparison away from the extended, moderate-gradient magnetized band in favor of the short, high-gradient unmagnetized caustic.

\phantomsection
\label{app:sens:settling}
\emph{Band evolution and temporal sampling.} The baseline history contains $24$ field snapshots.
Re-evaluating the $R=4\;\mu$m criterion with $F_r$ interpolated from its staggered radial grid reproduces the archived band edges.
From $z=27.9$~mm to the endpoint, the first sample outside the rear edge fails $F_r<0$, while the first sample beyond the front edge fails $E_z>0.10\,E_0$.
This identifies the field boundaries responsible for the late contraction.

Within the early comparison window ($9.9$--$19.5$~mm), the baseline field archive contains two snapshots, at $12.0$ and $15.9$~mm.
Their band centers give $-282.5\pm14.5\;\mu$m.
The fourteen snapshots from $43.8$ to $59.9$~mm give $-223.0\pm1.2\;\mu$m; both spreads are temporal standard deviations with divisor $n$.
The difference of the window means is $59.5\;\mu$m.
Figure~\ref{fig:bridge} uses these same two and fourteen snapshots for its profiles and band edges.
The separate field-strength scans use the five-snapshot density observable specified in Appendix~\ref{app:sens:selection}.

\subsection{Image extraction and positioning-law sensitivity}
\label{app:sens:statistics}
\label{app:sens:selection}
\label{sec:law:observable}
\label{sec:law:uncertainty}

Electron density profiles at the innermost physical sample ($r=2\;\mu$m) are averaged over five snapshots at $t\ge33$~ps and smoothed over $11\;\mu$m, excluding the region within $40\;\mu$m behind the driver.
The calibration and electron-driver analyses identify the first-image candidates near the positioning-law prediction.
For the electron-driver scan, the nearest peak is selected at each $\Om$ and the common anchor is refitted iteratively until the assignment is self-consistent.
The resulting positions and residuals are listed in Table~\ref{tab:lawres}.
Freeing both slope and anchor gives $\lambda=359.7\;\mu$m and an rms residual of $9.39\;\mu$m.

\begin{figure}
\includegraphics[width=\columnwidth]{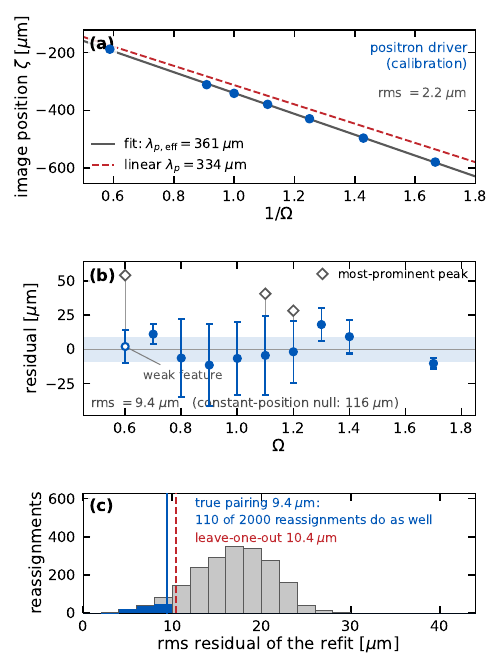}
\caption{\label{fig:law:selection}Peak-association checks for the positioning law.
(a) Positron-driver calibration.
(b) Electron-driver residuals using the calibrated wavelength and fitted anchor; the shaded band denotes the rms residual and error bars show the per-run temporal standard deviation over five snapshots.
Diamonds mark the three alternative most-prominent peaks, evaluated against the same prediction line; the open circle marks the weak $\Om=0.6$ feature.
(c) Residual distribution for $2000$ magnetic-field reassignments using the same multistart search as the physical pairing.
The solid line marks the physical-pairing residual and the dashed line the nearest-peak leave-one-out residual.}
\end{figure}

Figure~\ref{fig:law:selection} compares the fitted positions with alternative peak choices and randomized assignments.
The topographically most prominent peak, selected without consulting the law, agrees with the nearest-peak choice at seven of ten field strengths.
Fitting these independent selections gives a $21.5\;\mu$m rms residual, compared with $110.8\;\mu$m for a constant-position model, an improvement by a factor of $5.2$.
At $\Om=1.1$ and $1.2$, the two rules select different summits of the same density structure; the law-associated feature at $\Om=0.6$ is weak.
Removing candidates with prominence below $0.005\,\n0$ reduces the noise peaks without changing the primary selections.

The permutation test randomly reassigns the ten field strengths to the ten candidate-peak sets and repeats the fit.
The true assignment and each permutation receive the same search over $100$ initial anchors, retaining the smallest residual.
Among $2000$ reassignments, $110$ match or improve on the physical pairing's $9.4\;\mu$m rms residual; the null median is $17.0\;\mu$m and $p\simeq0.055$.
This comparison quantifies the contribution of candidate selection to a small residual.

In leave-one-out tests, the anchor is fitted on nine runs and the withheld prediction is scored against the nearest candidate peak in the remaining run.
The anchor changes by at most $2.0\;\mu$m, and the withheld rms residual is $10.4\;\mu$m.
This tests the stability of the same peak-association procedure; the withheld peak is still chosen relative to the prediction.

The error bars in Fig.~\ref{fig:law:selection}(b) show temporal variability: the sample standard deviation of the peak positions across five snapshots, using the fitted phase $\zeta_c-\lambda_{p,\mathrm{eff}}/\Om$ as the fixed reference for the nearest-peak choice in each snapshot.
These spreads include wake evolution and changes in peak association.
They describe the underlying snapshots, rather than an uncertainty assigned to the peak of the time-averaged profile.

\begin{table}
\caption{\label{tab:lawres}Positioning-law data and residuals relative to $\zeta_1=\zeta_c-\lambda_{p,\mathrm{eff}}/\Om$. \emph{Snapshot std} is the temporal sample standard deviation of peak position over five snapshots at $t\ge33$~ps: the four scan-window snapshots and $t=35.6$~ps.
The per-snapshot selection uses the fitted phase as a fixed reference (Appendix~\ref{app:sens:selection}).}
\begin{ruledtabular}
\begin{tabular}{cccc}
$\Om$ & peak $\zeta$ & snapshot std & residual \\
 & [$\mu$m] & [$\mu$m] & [$\mu$m] \\
\hline
$0.6$ & $-551$ & $12.0$ & $+2.1$ \\
$0.7$ & $-456$ & $7.2$ & $+11.1$ \\
$0.8$ & $-409$ & $28.3$ & $-6.3$ \\
$0.9$ & $-364$ & $30.0$ & $-11.5$ \\
$1.0$ & $-319$ & $26.5$ & $-6.6$ \\
$1.1$ & $-284$ & $29.0$ & $-4.4$ \\
$1.2$ & $-254$ & $22.5$ & $-1.7$ \\
$1.3$ & $-211$ & $12.2$ & $+18.1$ \\
$1.4$ & $-200$ & $12.3$ & $+9.3$ \\
$1.7$ & $-174$ & $3.7$ & $-10.2$ \\
\end{tabular}
\end{ruledtabular}
\end{table}

\subsection{Tracer sampling and perigee extraction}
\label{app:sens:tracer}
\label{sec:tracer:setup}

We introduce $2800$ negligibly loading plasma-electron tracers, with $400$ at each launch radius $r_0=2$, $6$, $10$, $14$, $18$, $22$, and $26\;\mu$m.
The cadence is $0.0504\,T_c$, and elapsed time is referenced to the driver centroid's passage across each particle's initial position.
We apply a five-sample running median to $r(\tau)$ and select the first local minimum in $0.25<\tau<2.0$~ps whose prominence is at least the larger of $2\;\mu$m and $20\%$ of the radial excursion within that window.
If none qualifies, an interior global minimum in $0.25<\tau<1.7$~ps is used.
The second perigee is the next qualifying filtered minimum sought $0.4$--$1.8$~ps after the first.
The unmagnetized control uses the same launch ensemble; its minimum radii are evaluated over $0.25<\tau<1.6$~ps.

The perigee positions use the centroid-passage origin, $\zeta=0$.
The median first-return position gives $-\zeta(\tau^*)=262.5\;\mu$m.
Table~\ref{tab:tracer} uses this centroid-referenced convention.

Table~\ref{tab:tracer} reports the six bins at $r_0=2$--$22\;\mu$m; ensemble-wide statistics include the additional $26\;\mu$m bin.
Estimator (i) reads the filtered radius at each particle's selected perigee time.
Estimator (ii) reads the unfiltered radius at that same time, without independently searching for a new minimum.
Estimator (iii) reads the pooled running-median radial curve at the bin-median perigee time.
The pooled curve uses a $0.02$~ps time grid and a $\pm0.03$~ps neighborhood.
These definitions preserve the launch-radius ordering while differing most near the axis.

\emph{Canonical angular momentum.} For $\ell=rp_\theta-\Om r^2/2$, using the normalization of Sec.~\ref{sec:theory:invariant}, median relative drifts over $1.5\,T_c$ are $2.4\%$--$4.0\%$ in the $r_0=14$, $18$, and $22\;\mu$m bins.

\phantomsection
\label{app:impl:psi}
\emph{Pseudo-potential reconstruction.} We integrate $\partial\psi/\partial\zeta=-E_z$ from the undisturbed region using field snapshots from the tracer run itself, and interpolate in time and space onto each trajectory.
For $C=\gamma-p_z-1-\psi_{\rm field}$, the temporal standard deviation along an orbit has an ensemble median of $0.0048$ (central $90\%$ range $0.0030$--$0.0060$, $2800$ orbits).
This quantifies the variation of the quasi-static invariant along each return.
The orbit means of $\psi_{\rm field}$, weighted by $\mathrm{d}t/\gamma$, have a median of $0.0754$ and a central $90\%$ range of $0.0448$--$0.1108$.
The weighting corresponds to the magnetic contribution to gyrophase advance.
The observed momentum rotation has a median of $0.86$ turns (central $90\%$: $0.71$--$0.96$), and the ratio of the integrated magnetic rotation to the observed rotation has a median of $0.92$ ($0.82$--$0.99$).

\begin{table}
\caption{\label{tab:tracer}Tracer statistics by launch radius, $400$ particles per bin, cadence $0.0504\,T_c$.  $\tau^*$ is the first perigee time; columns (i) and (ii) give bin medians of the filtered and unfiltered radii at the same per-particle perigee times; column (iii) samples the pooled running-median curve at the bin-median perigee time.  $\Delta\zeta_1$ and $\Delta\zeta_2$ denote $-\zeta$ at the first and second perigee, referenced to the driver centroid; both return times use estimator (i).}
\begin{ruledtabular}
\begin{tabular}{ccccccc}
$r_0$ & $\tau^*/T_c$ & \multicolumn{3}{c}{perigee radius [$\mu$m]} &
$\Delta\zeta_1$ & $\Delta\zeta_2$ \\
$[\mu$m$]$ & & (i) & (ii) & (iii) & [$\mu$m] & [$\mu$m] \\
\hline
$2$ & $0.96$ & $5.2$ & $1.1$ & $2.4$ & $282$ & $614$ \\
$6$ & $1.10$ & $6.6$ & $1.7$ & $2.5$ & $311$ & $629$ \\
$10$ & $1.06$ & $7.1$ & $3.3$ & $3.6$ & $298$ & $624$ \\
$14$ & $0.94$ & $8.7$ & $5.8$ & $6.3$ & $263$ & $603$ \\
$18$ & $0.81$ & $10.5$ & $7.4$ & $7.9$ & $229$ & $502$ \\
$22$ & $0.74$ & $12.1$ & $7.7$ & $8.0$ & $214$ & $446$ \\
\end{tabular}
\end{ruledtabular}
\end{table}

\subsection{Witness transport and resolution dependence}
\label{app:sens:resolution}

\phantomsection
\label{app:sens:finegrid}
\emph{Weak-force core and aperture limits.} For the eleven cadence snapshots in Fig.~\ref{fig:aperture}, the fine-grid median lengths at $R=4$, $6$, $8$, $12$, and $16\;\mu$m are $24$, $23$, $21$, $14$, and $0\;\mu$m.
A band containing the witness appears in four snapshots, while the criterion applied to snapshot-median fields gives zero length.
The registered baseline gives $L(4)=20$--$24\;\mu$m over its eleven matched-interval snapshots, with median lengths $21$, $21$, $21$, $6$, and $0\;\mu$m at the same radii.

For the force-profile comparison, fields are averaged across $\zeta\in[-235,-219]\;\mu$m and then across twelve fine-grid or thirteen baseline snapshots.
The fine-grid mean force inside $6\;\mu$m is at most $1.8$ standard errors from zero, whereas the focusing ring exceeds ten standard errors in magnitude from $r=9.5\;\mu$m outward; the errors describe snapshot variation.
Interpolating the baseline transverse field to physical $r=4\;\mu$m gives $F_r=-0.0523\,E_0$.

\phantomsection
\label{app:sens:branch}
\emph{Resolution dependence of the energy gain.} Table~\ref{tab:convergence} uses the final series outputs at $199.92$, $200.00$, $200.00$, and $200.26$~ps for the baseline and three WarpX grids.
For the $4$ and $2\;\mu$m WarpX runs, captured fractions stored relative to the initial in-aperture count are converted to injected-charge fractions using the factor $0.99095$ from the $1\;\mu$m run.
The WarpX emittances average the two projected transverse planes.

Interpolating each WarpX gain history to $36$ and $99$~ps gives $\Delta\gamma=(30.1,83.3)$, $(32.4,84.8)$, and $(33.6,88.7)$ on the $4$, $2$, and $1\;\mu$m grids.
At $99$~ps, the baseline has gained $152.9$, giving a baseline--fine-grid gap of $64.2$, compared with approximately $100$ at the endpoint.
Linear fits over the final $30$~ps of the baseline and $40$~ps of the finest WarpX run give $1.88$ and $2.37$~GeV/m, respectively.
The early cross-code gap and late refinement dependence therefore enter different portions of the energy integral.

\phantomsection
\label{app:impl:emittance}
\emph{Mechanical emittance and slices.} The $1\;\mu$m WarpX run records the four-dimensional mechanical beam matrix.
At $200.26$~ps, its eigen-emittances are $82.92$ and $82.44$~mm\,mrad, compared with projected values of $82.50$ and $82.88$~mm\,mrad.
The magnitude of $\langle xu_y-yu_x\rangle$ is $0.25\;\mu$m.
The six equal-charge longitudinal slices give $x$-plane emittances from $64.45$ to $81.50$~mm\,mrad, with median $72.90$, and a median relative energy spread of $0.92\%$.

\emph{Low-emittance resolution control.} We average the two projected plane emittances and interpolate the $1\;\mu$m run onto the actual output times of the $0.5\;\mu$m radial-grid control.
The maximum relative difference is $15.1\%$ over their common interval.
The control ends at $87.63$~ps ($26.27$~mm), by which time the $1\;\mu$m run has accumulated $70.6\%$ of its full-stage emittance increase.
The $38.0$~mm\,mrad final value belongs to the longer $1\;\mu$m run.

\subsection{Beam loading and witness-profile shaping}
\label{app:sens:loading}

The charge scan keeps the witness macroparticle count fixed.
The loading-field statistics use on-axis fields smoothed over $11\;\mu$m, averaged over $t\ge140$~ps, and then over $\zeta\in[-235,-219]\;\mu$m.
The mean-field response through $400$~pC has a slope of approximately $-1.0\times10^{-3}\,E_0$ per pC.

On the baseline grid, the core spring estimator evaluated at $r=2\;\mu$m yields $(28.9\pm3.2)\times10^3$~m$^{-1}$ for the $0.1$~pC witness (standard error over $14$ snapshots).
The values are $30.5$, $30.0$, $28.8$, $27.8$, and $30.8\times10^3$~m$^{-1}$ across the $25$ to $400$~pC scan.
At $800$~pC, it rises to $(47.5\pm4.2)\times10^3$~m$^{-1}$ (standard error over $11$ snapshots).

The extraction efficiency is $\eta_{\rm extr}=\Delta U_w/(-\Delta U_d)$, using the kinetic-energy changes of the entire witness and driver populations, including witness particles outside $\rcap$.
Equation~(\ref{eq:lumi}) combines this efficiency and the injected population $N=Q/e$ with the captured-ensemble emittance.
The loading scan uses the common $99$~ps epoch; its best sampled value is $\tilde L_P\simeq0.07$.

For the $0.1$ and $200$~pC runs at $99$~ps, the captured-ensemble emittance is $\sqrt{\varepsilon_{nx}\varepsilon_{ny}}$.
The $0.1$~pC witness gives $\eta_{\rm extr}=1.44\times10^{-4}$, $\sqrt{\varepsilon_{nx}\varepsilon_{ny}}=65.2$~mm\,mrad, and $\tilde L_P=4.88\times10^{-8}$.
The $200$~pC value, $0.0733$, is larger by a factor of $1.50\times10^6$.
For both runs, the particle and energy diagnostics are taken at $98.959$ and $99.058$~ps, respectively.

The electron-motion parameter of Ref.~\cite{Cao2024}, $k_e\sigma_z=k_p\sigma_z\sqrt{n_b/(2\n0)}$, reaches unity near $n_b=350\,\n0$ (about $1.7$~nC), beyond the scanned charge range.
The proposed magnetization threshold $n_b<2\Om^2\n0\simeq2.4\,\n0$ would instead predict degradation near $25$~pC, where emittance changes by only $0.2\%$.
This threshold does not account for the observed charge tolerance.

\phantomsection
\label{app:sens:kernel}
\emph{Wake response for profile shaping.} The shaped witness profile tested in Sec.~\ref{sec:loading:shaping} was derived from a linearized wake response kernel $K(\zeta)$, assuming a linear scaling up to $200$~pC.

The design-profile-weighted statistics use the injected profile, with each shaped slice represented by its uniform $2\;\mu$m support: $w(\zeta)=\sum_j(w_j/2)\,\mathbf{1}_{[\zeta_j-1,\zeta_j+1)}(\zeta)$, where lengths are in micrometers.
The on-axis field is smoothed over $11\;\mu$m and averaged over the eleven snapshots at $t\ge140$~ps before its spatial mean and standard deviation are evaluated with these weights.
The shaped and Gaussian means are $0.2852\,E_0$ and $0.2024\,E_0$, with relative standard deviations of $12.95\%$ and $12.73\%$.

\subsection{Azimuthal and entrance-state diagnostics}
\label{app:sens:stability}

The dipole test uses the working-bucket window, while the quadrupole/sextupole test uses a broader wake window; the full-3D analysis separately samples the working bucket and a deeper window.
Quiet mirror-symmetric initialization gives $A_1\sim10^{-11}$ through $150$~ps.
For the seeded dipole pair, the $10$--$100$~ps growth factors are $8.72$ with magnetization and $4.45$ without it.
For the $2\%$ higher-mode seeds, $A_2$ changes from $0.059$ to $0.109$ and $A_3$ from $0.029$ to $0.053$ between $35$ and $99$~ps.
The interval before approximately $45$~ps precedes appreciable driver-seed growth; the later evolution samples a nonlinear response.
In 3D, $A_2\le0.0035$ and $A_3\le0.0075$ in both analysis windows.

\phantomsection\label{app:sens:magnet}
\emph{Entrance-state comparison.} The driver is initialized with $v_\phi=\Om k_pr\,c/(2\gamma_d)$.
Its head-radius ratio to the matched-time $p_\phi=0$ reference is $0.9864\pm0.0032$ (standard error over ten samples), compared with the paraxial vacuum estimate $0.9886$.
This head-radius reference has the same driver and field but a different witness charge from the image-position baseline.
The density-image peak shifts backward by $10\;\mu$m.
Removing the last snapshot from both image comparisons, to reduce the effect of offset endpoint epochs, gives a $7\;\mu$m image shift.

For a cold, initially parallel driver, the paraxial vacuum model gives $k_L=\Om k_p/(2\gamma_d)\simeq10$~m$^{-1}$ and predicts a radius ratio of approximately $0.81$ after $60$~mm, beyond the $18.9$~mm covered by the entrance-state simulation.
The witness small-kick estimate of the relative projected-emittance change, $\tfrac12(0.072/7.14)^2\simeq5\times10^{-5}$, applies to the assumed fringe model; evaluating a specific exit configuration requires extracted-beam tracking.

\section{Additional operating limits}
\label{app:limits}

Complementing the beam-placement and tolerance tests in the main text, this appendix examines how the initial plasma temperature and the finite transverse width of the plasma column affect the formation of the magnetized wake structure.

\subsection{Plasma temperature}
\label{sec:stability:temperature}

The baseline simulations assume an initially cold plasma.
A finite electron temperature spreads the return phases and transverse orbits.
We estimate these effects using the longitudinal spread $\Delta\zeta\approx(\lp/\Om)(v_{\rm th}/c)$ and the thermal Larmor radius, using the one-dimensional rms thermal speed $v_{\rm th}=c\sqrt{T_e/(m_ec^2)}$, with $T_e$ expressed in energy units.
At electron temperatures of $10$, $100$, and $1000$~eV, the longitudinal thermal spreads are $1.3$, $4.2$, and $13.4\;\mu$m, respectively, compared with a typical settled band length of roughly $22\;\mu$m.
The corresponding thermal Larmor radii are $0.21$, $0.68$, and $2.1\;\mu$m.
These scales suggest increasing broadening of the gyro images as temperature rises.

Formation tests compare the early peak density of the returning electrons with the corresponding cold-plasma reference.
At $100$~eV and $40$~ps, the magnetized peak remains close to its cold value, with a warm-to-cold ratio of $1.04$, while the unmagnetized caustic falls to $0.55$ and develops a tail.
At $30$~ps, the corresponding ratios are $0.72$ and $0.65$.
The magnetized image undergoes a smaller relative peak-density reduction at both epochs, although the separation changes as the structures evolve.
At $1000$~eV, the magnetized peak at $30$~ps falls to about half its cold-reference value, and the image broadens during formation.

\subsection{Plasma width}
\label{app:limits:width}

Separate diagnostic tests at $\n0=10^{13}\;\mathrm{cm^{-3}}$ and $\Om=0.9857$ compare finite flat-top plasma radii $R_f=1.19\,c/\wp$ and $0.37\,c/\wp$ with a full-width reference at the common epoch of $313.3$~ps.

The wider column retains a weakened oscillatory wake.
At $R_f=0.37\,c/\wp$, the autocorrelation peak of the on-axis accelerating field drops from $0.35$ to $0.03$, while driver energy loss falls to $7.5\%$ of the reference value.
The on-axis field remains single-signed over the final six skin depths of the analysis window, with its magnitude increasing from $0.15$ to $0.29\,E_0$.
Together, these diagnostics indicate a loss of the oscillatory wake despite the larger field magnitude.

The loss of oscillatory structure in the narrow flat-top column highlights the role of plasma width in the collective return response.
At fixed $\Om$ and normalized radius $k_pR_f$, lowering $\n0$ reduces the required magnetic field as $B\propto\sqrt{\n0}$ but increases the physical plasma radius as $R_f\propto\n0^{-1/2}$.

\clearpage
\bibliography{references}

\begin{thebibliography}{48}%
\makeatletter
\providecommand \@ifxundefined [1]{%
 \@ifx{#1\undefined}
}%
\providecommand \@ifnum [1]{%
 \ifnum #1\expandafter \@firstoftwo
 \else \expandafter \@secondoftwo
 \fi
}%
\providecommand \@ifx [1]{%
 \ifx #1\expandafter \@firstoftwo
 \else \expandafter \@secondoftwo
 \fi
}%
\providecommand \natexlab [1]{#1}%
\providecommand \enquote  [1]{``#1''}%
\providecommand \bibnamefont  [1]{#1}%
\providecommand \bibfnamefont [1]{#1}%
\providecommand \citenamefont [1]{#1}%
\providecommand \href@noop [0]{\@secondoftwo}%
\providecommand \href [0]{\begingroup \@sanitize@url \@href}%
\providecommand \@href[1]{\@@startlink{#1}\@@href}%
\providecommand \@@href[1]{\endgroup#1\@@endlink}%
\providecommand \@sanitize@url [0]{\catcode `\\12\catcode `\$12\catcode
  `\&12\catcode `\#12\catcode `\^12\catcode `\_12\catcode `\%12\relax}%
\providecommand \@@startlink[1]{}%
\providecommand \@@endlink[0]{}%
\providecommand \url  [0]{\begingroup\@sanitize@url \@url }%
\providecommand \@url [1]{\endgroup\@href {#1}{\urlprefix }}%
\providecommand \urlprefix  [0]{URL }%
\providecommand \Eprint [0]{\href }%
\providecommand \doibase [0]{https://doi.org/}%
\providecommand \selectlanguage [0]{\@gobble}%
\providecommand \bibinfo  [0]{\@secondoftwo}%
\providecommand \bibfield  [0]{\@secondoftwo}%
\providecommand \translation [1]{[#1]}%
\providecommand \BibitemOpen [0]{}%
\providecommand \bibitemStop [0]{}%
\providecommand \bibitemNoStop [0]{.\EOS\space}%
\providecommand \EOS [0]{\spacefactor3000\relax}%
\providecommand \BibitemShut  [1]{\csname bibitem#1\endcsname}%
\let\auto@bib@innerbib\@empty
\bibitem [{\citenamefont {Tajima}\ and\ \citenamefont
  {Dawson}(1979)}]{Tajima1979}%
  \BibitemOpen
  \bibfield  {author} {\bibinfo {author} {\bibfnamefont {T.}~\bibnamefont
  {Tajima}}\ and\ \bibinfo {author} {\bibfnamefont {J.~M.}\ \bibnamefont
  {Dawson}},\ }\bibfield  {title} {\bibinfo {title} {Laser electron
  accelerator},\ }\href@noop {} {\bibfield  {journal} {\bibinfo  {journal}
  {Phys. Rev. Lett.}\ }\textbf {\bibinfo {volume} {43}},\ \bibinfo {pages}
  {267} (\bibinfo {year} {1979})}\BibitemShut {NoStop}%
\bibitem [{\citenamefont {Chen}\ \emph {et~al.}(1985)\citenamefont {Chen},
  \citenamefont {Dawson}, \citenamefont {Huff},\ and\ \citenamefont
  {Katsouleas}}]{Chen1985}%
  \BibitemOpen
  \bibfield  {author} {\bibinfo {author} {\bibfnamefont {P.}~\bibnamefont
  {Chen}}, \bibinfo {author} {\bibfnamefont {J.~M.}\ \bibnamefont {Dawson}},
  \bibinfo {author} {\bibfnamefont {R.~W.}\ \bibnamefont {Huff}},\ and\
  \bibinfo {author} {\bibfnamefont {T.}~\bibnamefont {Katsouleas}},\ }\bibfield
   {title} {\bibinfo {title} {Acceleration of electrons by the interaction of a
  bunched electron beam with a plasma},\ }\href@noop {} {\bibfield  {journal}
  {\bibinfo  {journal} {Phys. Rev. Lett.}\ }\textbf {\bibinfo {volume} {54}},\
  \bibinfo {pages} {693} (\bibinfo {year} {1985})}\BibitemShut {NoStop}%
\bibitem [{\citenamefont {Blumenfeld}\ \emph {et~al.}(2007)\citenamefont
  {Blumenfeld} \emph {et~al.}}]{Blumenfeld2007}%
  \BibitemOpen
  \bibfield  {author} {\bibinfo {author} {\bibfnamefont {I.}~\bibnamefont
  {Blumenfeld}} \emph {et~al.},\ }\bibfield  {title} {\bibinfo {title} {Energy
  doubling of 42 {GeV} electrons in a metre-scale plasma wakefield
  accelerator},\ }\href@noop {} {\bibfield  {journal} {\bibinfo  {journal}
  {Nature}\ }\textbf {\bibinfo {volume} {445}},\ \bibinfo {pages} {741}
  (\bibinfo {year} {2007})}\BibitemShut {NoStop}%
\bibitem [{\citenamefont {Litos}\ \emph {et~al.}(2014)\citenamefont {Litos}
  \emph {et~al.}}]{Litos2014}%
  \BibitemOpen
  \bibfield  {author} {\bibinfo {author} {\bibfnamefont {M.}~\bibnamefont
  {Litos}} \emph {et~al.},\ }\bibfield  {title} {\bibinfo {title}
  {High-efficiency acceleration of an electron beam in a plasma wakefield
  accelerator},\ }\href@noop {} {\bibfield  {journal} {\bibinfo  {journal}
  {Nature}\ }\textbf {\bibinfo {volume} {515}},\ \bibinfo {pages} {92}
  (\bibinfo {year} {2014})}\BibitemShut {NoStop}%
\bibitem [{\citenamefont {Lindstr{\o}m}\ \emph {et~al.}(2024)\citenamefont
  {Lindstr{\o}m} \emph {et~al.}}]{Lindstrom2024}%
  \BibitemOpen
  \bibfield  {author} {\bibinfo {author} {\bibfnamefont {C.~A.}\ \bibnamefont
  {Lindstr{\o}m}} \emph {et~al.},\ }\bibfield  {title} {\bibinfo {title}
  {Emittance preservation in a plasma-wakefield accelerator},\ }\href@noop {}
  {\bibfield  {journal} {\bibinfo  {journal} {Nat. Commun.}\ }\textbf {\bibinfo
  {volume} {15}},\ \bibinfo {pages} {6097} (\bibinfo {year}
  {2024})}\BibitemShut {NoStop}%
\bibitem [{\citenamefont {Cao}\ \emph {et~al.}(2024)\citenamefont {Cao},
  \citenamefont {Lindstr{\o}m}, \citenamefont {Adli}, \citenamefont {Corde},\
  and\ \citenamefont {Gessner}}]{Cao2024}%
  \BibitemOpen
  \bibfield  {author} {\bibinfo {author} {\bibfnamefont {G.~J.}\ \bibnamefont
  {Cao}}, \bibinfo {author} {\bibfnamefont {C.~A.}\ \bibnamefont
  {Lindstr{\o}m}}, \bibinfo {author} {\bibfnamefont {E.}~\bibnamefont {Adli}},
  \bibinfo {author} {\bibfnamefont {S.}~\bibnamefont {Corde}},\ and\ \bibinfo
  {author} {\bibfnamefont {S.}~\bibnamefont {Gessner}},\ }\bibfield  {title}
  {\bibinfo {title} {Positron acceleration in plasma wakefields},\ }\href@noop
  {} {\bibfield  {journal} {\bibinfo  {journal} {Phys. Rev. Accel. Beams}\
  }\textbf {\bibinfo {volume} {27}},\ \bibinfo {pages} {034801} (\bibinfo
  {year} {2024})}\BibitemShut {NoStop}%
\bibitem [{\citenamefont {Joshi}\ \emph {et~al.}(2025)\citenamefont {Joshi},
  \citenamefont {Mori},\ and\ \citenamefont {Hogan}}]{Joshi2025}%
  \BibitemOpen
  \bibfield  {author} {\bibinfo {author} {\bibfnamefont {C.}~\bibnamefont
  {Joshi}}, \bibinfo {author} {\bibfnamefont {W.~B.}\ \bibnamefont {Mori}},\
  and\ \bibinfo {author} {\bibfnamefont {M.~J.}\ \bibnamefont {Hogan}},\
  }\bibfield  {title} {\bibinfo {title} {The positron arm of a plasma-based
  linear collider},\ }\href@noop {} {\bibfield  {journal} {\bibinfo  {journal}
  {Nat. Phys.}\ }\textbf {\bibinfo {volume} {21}},\ \bibinfo {pages} {885}
  (\bibinfo {year} {2025})}\BibitemShut {NoStop}%
\bibitem [{\citenamefont {Chen}\ and\ \citenamefont {Liu}(2026)}]{ChenLiu2026}%
  \BibitemOpen
  \bibfield  {author} {\bibinfo {author} {\bibfnamefont {P.}~\bibnamefont
  {Chen}}\ and\ \bibinfo {author} {\bibfnamefont {Y.-K.}\ \bibnamefont {Liu}},\
  }\bibfield  {title} {\bibinfo {title} {Plasma wakefield: from accelerators to
  black holes},\ }\href {https://doi.org/10.1007/s41614-026-00217-x} {\bibfield
   {journal} {\bibinfo  {journal} {Reviews of Modern Plasma Physics}\ }\textbf
  {\bibinfo {volume} {10}},\ \bibinfo {pages} {7} (\bibinfo {year}
  {2026})}\BibitemShut {NoStop}%
\bibitem [{\citenamefont {Rosenzweig}\ \emph {et~al.}(1991)\citenamefont
  {Rosenzweig}, \citenamefont {Breizman}, \citenamefont {Katsouleas},\ and\
  \citenamefont {Su}}]{Rosenzweig1991}%
  \BibitemOpen
  \bibfield  {author} {\bibinfo {author} {\bibfnamefont {J.~B.}\ \bibnamefont
  {Rosenzweig}}, \bibinfo {author} {\bibfnamefont {B.}~\bibnamefont
  {Breizman}}, \bibinfo {author} {\bibfnamefont {T.}~\bibnamefont
  {Katsouleas}},\ and\ \bibinfo {author} {\bibfnamefont {J.~J.}\ \bibnamefont
  {Su}},\ }\bibfield  {title} {\bibinfo {title} {Acceleration and focusing of
  electrons in two-dimensional nonlinear plasma wake fields},\ }\href@noop {}
  {\bibfield  {journal} {\bibinfo  {journal} {Phys. Rev. A}\ }\textbf {\bibinfo
  {volume} {44}},\ \bibinfo {pages} {R6189} (\bibinfo {year}
  {1991})}\BibitemShut {NoStop}%
\bibitem [{\citenamefont {Lu}\ \emph {et~al.}(2006)\citenamefont {Lu},
  \citenamefont {Huang}, \citenamefont {Zhou}, \citenamefont {Mori},\ and\
  \citenamefont {Katsouleas}}]{Lu2006}%
  \BibitemOpen
  \bibfield  {author} {\bibinfo {author} {\bibfnamefont {W.}~\bibnamefont
  {Lu}}, \bibinfo {author} {\bibfnamefont {C.}~\bibnamefont {Huang}}, \bibinfo
  {author} {\bibfnamefont {M.}~\bibnamefont {Zhou}}, \bibinfo {author}
  {\bibfnamefont {W.~B.}\ \bibnamefont {Mori}},\ and\ \bibinfo {author}
  {\bibfnamefont {T.}~\bibnamefont {Katsouleas}},\ }\bibfield  {title}
  {\bibinfo {title} {Nonlinear theory for relativistic plasma wakefields in the
  blowout regime},\ }\href@noop {} {\bibfield  {journal} {\bibinfo  {journal}
  {Phys. Rev. Lett.}\ }\textbf {\bibinfo {volume} {96}},\ \bibinfo {pages}
  {165002} (\bibinfo {year} {2006})}\BibitemShut {NoStop}%
\bibitem [{\citenamefont {Lotov}(2007)}]{Lotov2007}%
  \BibitemOpen
  \bibfield  {author} {\bibinfo {author} {\bibfnamefont {K.~V.}\ \bibnamefont
  {Lotov}},\ }\bibfield  {title} {\bibinfo {title} {Acceleration of positrons
  by electron beam-driven wakefields in a plasma},\ }\href@noop {} {\bibfield
  {journal} {\bibinfo  {journal} {Phys. Plasmas}\ }\textbf {\bibinfo {volume}
  {14}},\ \bibinfo {pages} {023101} (\bibinfo {year} {2007})}\BibitemShut
  {NoStop}%
\bibitem [{\citenamefont {Gessner}\ \emph {et~al.}(2016)\citenamefont {Gessner}
  \emph {et~al.}}]{Gessner2016}%
  \BibitemOpen
  \bibfield  {author} {\bibinfo {author} {\bibfnamefont {S.}~\bibnamefont
  {Gessner}} \emph {et~al.},\ }\bibfield  {title} {\bibinfo {title}
  {Demonstration of a positron beam-driven hollow channel plasma wakefield
  accelerator},\ }\href@noop {} {\bibfield  {journal} {\bibinfo  {journal}
  {Nat. Commun.}\ }\textbf {\bibinfo {volume} {7}},\ \bibinfo {pages} {11785}
  (\bibinfo {year} {2016})}\BibitemShut {NoStop}%
\bibitem [{\citenamefont {Lindstr{\o}m}\ \emph {et~al.}(2018)\citenamefont
  {Lindstr{\o}m} \emph {et~al.}}]{Lindstrom2018}%
  \BibitemOpen
  \bibfield  {author} {\bibinfo {author} {\bibfnamefont {C.~A.}\ \bibnamefont
  {Lindstr{\o}m}} \emph {et~al.},\ }\bibfield  {title} {\bibinfo {title}
  {Measurement of transverse wakefields induced by a misaligned positron bunch
  in a hollow channel plasma accelerator},\ }\href
  {https://doi.org/10.1103/PhysRevLett.120.124802} {\bibfield  {journal}
  {\bibinfo  {journal} {Phys. Rev. Lett.}\ }\textbf {\bibinfo {volume} {120}},\
  \bibinfo {pages} {124802} (\bibinfo {year} {2018})}\BibitemShut {NoStop}%
\bibitem [{\citenamefont {Silva}\ \emph {et~al.}(2021)\citenamefont {Silva}
  \emph {et~al.}}]{Silva2021}%
  \BibitemOpen
  \bibfield  {author} {\bibinfo {author} {\bibfnamefont {T.}~\bibnamefont
  {Silva}} \emph {et~al.},\ }\bibfield  {title} {\bibinfo {title} {Stable
  positron acceleration in thin, warm, hollow plasma channels},\ }\href@noop {}
  {\bibfield  {journal} {\bibinfo  {journal} {Phys. Rev. Lett.}\ }\textbf
  {\bibinfo {volume} {127}},\ \bibinfo {pages} {104801} (\bibinfo {year}
  {2021})}\BibitemShut {NoStop}%
\bibitem [{\citenamefont {Diederichs}\ \emph {et~al.}(2019)\citenamefont
  {Diederichs}, \citenamefont {Mehrling}, \citenamefont {Benedetti},
  \citenamefont {Schroeder}, \citenamefont {Knetsch}, \citenamefont {Esarey},\
  and\ \citenamefont {Osterhoff}}]{Diederichs2019}%
  \BibitemOpen
  \bibfield  {author} {\bibinfo {author} {\bibfnamefont {S.}~\bibnamefont
  {Diederichs}}, \bibinfo {author} {\bibfnamefont {T.~J.}\ \bibnamefont
  {Mehrling}}, \bibinfo {author} {\bibfnamefont {C.}~\bibnamefont {Benedetti}},
  \bibinfo {author} {\bibfnamefont {C.~B.}\ \bibnamefont {Schroeder}}, \bibinfo
  {author} {\bibfnamefont {A.}~\bibnamefont {Knetsch}}, \bibinfo {author}
  {\bibfnamefont {E.}~\bibnamefont {Esarey}},\ and\ \bibinfo {author}
  {\bibfnamefont {J.}~\bibnamefont {Osterhoff}},\ }\bibfield  {title} {\bibinfo
  {title} {Positron transport and acceleration in beam-driven plasma wakefield
  accelerators using plasma columns},\ }\href@noop {} {\bibfield  {journal}
  {\bibinfo  {journal} {Phys. Rev. Accel. Beams}\ }\textbf {\bibinfo {volume}
  {22}},\ \bibinfo {pages} {081301} (\bibinfo {year} {2019})}\BibitemShut
  {NoStop}%
\bibitem [{\citenamefont {Diederichs}\ \emph {et~al.}(2022)\citenamefont
  {Diederichs}, \citenamefont {Benedetti}, \citenamefont {Th{\'e}venet},
  \citenamefont {Esarey}, \citenamefont {Osterhoff},\ and\ \citenamefont
  {Schroeder}}]{Diederichs2022}%
  \BibitemOpen
  \bibfield  {author} {\bibinfo {author} {\bibfnamefont {S.}~\bibnamefont
  {Diederichs}}, \bibinfo {author} {\bibfnamefont {C.}~\bibnamefont
  {Benedetti}}, \bibinfo {author} {\bibfnamefont {M.}~\bibnamefont
  {Th{\'e}venet}}, \bibinfo {author} {\bibfnamefont {E.}~\bibnamefont
  {Esarey}}, \bibinfo {author} {\bibfnamefont {J.}~\bibnamefont {Osterhoff}},\
  and\ \bibinfo {author} {\bibfnamefont {C.~B.}\ \bibnamefont {Schroeder}},\
  }\bibfield  {title} {\bibinfo {title} {Self-stabilizing positron acceleration
  in a plasma column},\ }\href@noop {} {\bibfield  {journal} {\bibinfo
  {journal} {Phys. Rev. Accel. Beams}\ }\textbf {\bibinfo {volume} {25}},\
  \bibinfo {pages} {091304} (\bibinfo {year} {2022})}\BibitemShut {NoStop}%
\bibitem [{\citenamefont {Corde}\ \emph {et~al.}(2015)\citenamefont {Corde}
  \emph {et~al.}}]{Corde2015}%
  \BibitemOpen
  \bibfield  {author} {\bibinfo {author} {\bibfnamefont {S.}~\bibnamefont
  {Corde}} \emph {et~al.},\ }\bibfield  {title} {\bibinfo {title}
  {Multi-gigaelectronvolt acceleration of positrons in a self-loaded plasma
  wakefield},\ }\href@noop {} {\bibfield  {journal} {\bibinfo  {journal}
  {Nature}\ }\textbf {\bibinfo {volume} {524}},\ \bibinfo {pages} {442}
  (\bibinfo {year} {2015})}\BibitemShut {NoStop}%
\bibitem [{\citenamefont {Zhou}\ \emph {et~al.}(2025)\citenamefont {Zhou},
  \citenamefont {Ding}, \citenamefont {An}, \citenamefont {Su}, \citenamefont
  {Hua}, \citenamefont {Li}, \citenamefont {Mori}, \citenamefont {Joshi},\ and\
  \citenamefont {Lu}}]{Zhou2025}%
  \BibitemOpen
  \bibfield  {author} {\bibinfo {author} {\bibfnamefont {S.}~\bibnamefont
  {Zhou}}, \bibinfo {author} {\bibfnamefont {S.}~\bibnamefont {Ding}}, \bibinfo
  {author} {\bibfnamefont {W.}~\bibnamefont {An}}, \bibinfo {author}
  {\bibfnamefont {Q.}~\bibnamefont {Su}}, \bibinfo {author} {\bibfnamefont
  {J.}~\bibnamefont {Hua}}, \bibinfo {author} {\bibfnamefont {F.}~\bibnamefont
  {Li}}, \bibinfo {author} {\bibfnamefont {W.~B.}\ \bibnamefont {Mori}},
  \bibinfo {author} {\bibfnamefont {C.}~\bibnamefont {Joshi}},\ and\ \bibinfo
  {author} {\bibfnamefont {W.}~\bibnamefont {Lu}},\ }\bibfield  {title}
  {\bibinfo {title} {Positron beam loading and acceleration in the blowout
  regime of a plasma wakefield accelerator},\ }\href@noop {} {\bibfield
  {journal} {\bibinfo  {journal} {Research}\ }\textbf {\bibinfo {volume} {8}},\
  \bibinfo {pages} {0878} (\bibinfo {year} {2025})}\BibitemShut {NoStop}%
\bibitem [{\citenamefont {Jain}\ \emph {et~al.}(2015)\citenamefont {Jain},
  \citenamefont {Antonsen},\ and\ \citenamefont {Palastro}}]{Jain2015}%
  \BibitemOpen
  \bibfield  {author} {\bibinfo {author} {\bibfnamefont {N.}~\bibnamefont
  {Jain}}, \bibinfo {author} {\bibfnamefont {T.~M.}\ \bibnamefont {Antonsen},
  \bibfnamefont {Jr.}},\ and\ \bibinfo {author} {\bibfnamefont {J.~P.}\
  \bibnamefont {Palastro}},\ }\bibfield  {title} {\bibinfo {title} {Positron
  acceleration by plasma wakefields driven by a hollow electron beam},\
  }\href@noop {} {\bibfield  {journal} {\bibinfo  {journal} {Phys. Rev. Lett.}\
  }\textbf {\bibinfo {volume} {115}},\ \bibinfo {pages} {195001} (\bibinfo
  {year} {2015})}\BibitemShut {NoStop}%
\bibitem [{\citenamefont {Vieira}\ and\ \citenamefont {Mendon{\c
  c}a}(2014)}]{Vieira2014}%
  \BibitemOpen
  \bibfield  {author} {\bibinfo {author} {\bibfnamefont {J.}~\bibnamefont
  {Vieira}}\ and\ \bibinfo {author} {\bibfnamefont {J.~T.}\ \bibnamefont
  {Mendon{\c c}a}},\ }\bibfield  {title} {\bibinfo {title} {Nonlinear laser
  driven donut wakefields for positron and electron acceleration},\ }\href@noop
  {} {\bibfield  {journal} {\bibinfo  {journal} {Phys. Rev. Lett.}\ }\textbf
  {\bibinfo {volume} {112}},\ \bibinfo {pages} {215001} (\bibinfo {year}
  {2014})}\BibitemShut {NoStop}%
\bibitem [{\citenamefont {Doche}\ \emph {et~al.}(2017)\citenamefont {Doche}
  \emph {et~al.}}]{Doche2017}%
  \BibitemOpen
  \bibfield  {author} {\bibinfo {author} {\bibfnamefont {A.}~\bibnamefont
  {Doche}} \emph {et~al.},\ }\bibfield  {title} {\bibinfo {title} {Acceleration
  of a trailing positron bunch in a plasma wakefield accelerator},\ }\href@noop
  {} {\bibfield  {journal} {\bibinfo  {journal} {Sci. Rep.}\ }\textbf {\bibinfo
  {volume} {7}},\ \bibinfo {pages} {14180} (\bibinfo {year}
  {2017})}\BibitemShut {NoStop}%
\bibitem [{\citenamefont {Xu}\ \emph {et~al.}(2020)\citenamefont {Xu},
  \citenamefont {Yi}, \citenamefont {Shen}, \citenamefont {Xu}, \citenamefont
  {Ji}, \citenamefont {Xu}, \citenamefont {Zhang}, \citenamefont {Li},\ and\
  \citenamefont {Xu}}]{Xu2020}%
  \BibitemOpen
  \bibfield  {author} {\bibinfo {author} {\bibfnamefont {Z.}~\bibnamefont
  {Xu}}, \bibinfo {author} {\bibfnamefont {L.}~\bibnamefont {Yi}}, \bibinfo
  {author} {\bibfnamefont {B.}~\bibnamefont {Shen}}, \bibinfo {author}
  {\bibfnamefont {J.}~\bibnamefont {Xu}}, \bibinfo {author} {\bibfnamefont
  {L.}~\bibnamefont {Ji}}, \bibinfo {author} {\bibfnamefont {T.}~\bibnamefont
  {Xu}}, \bibinfo {author} {\bibfnamefont {L.}~\bibnamefont {Zhang}}, \bibinfo
  {author} {\bibfnamefont {S.}~\bibnamefont {Li}},\ and\ \bibinfo {author}
  {\bibfnamefont {Z.}~\bibnamefont {Xu}},\ }\bibfield  {title} {\bibinfo
  {title} {Driving positron beam acceleration with coherent transition
  radiation},\ }\href@noop {} {\bibfield  {journal} {\bibinfo  {journal}
  {Commun. Phys.}\ }\textbf {\bibinfo {volume} {3}},\ \bibinfo {pages} {191}
  (\bibinfo {year} {2020})}\BibitemShut {NoStop}%
\bibitem [{\citenamefont {Su}\ \emph {et~al.}(1987)\citenamefont {Su},
  \citenamefont {Katsouleas}, \citenamefont {Dawson}, \citenamefont {Chen},
  \citenamefont {Jones},\ and\ \citenamefont {Keinigs}}]{Su1987}%
  \BibitemOpen
  \bibfield  {author} {\bibinfo {author} {\bibfnamefont {J.~J.}\ \bibnamefont
  {Su}}, \bibinfo {author} {\bibfnamefont {T.}~\bibnamefont {Katsouleas}},
  \bibinfo {author} {\bibfnamefont {J.~M.}\ \bibnamefont {Dawson}}, \bibinfo
  {author} {\bibfnamefont {P.}~\bibnamefont {Chen}}, \bibinfo {author}
  {\bibfnamefont {M.}~\bibnamefont {Jones}},\ and\ \bibinfo {author}
  {\bibfnamefont {R.}~\bibnamefont {Keinigs}},\ }\bibfield  {title} {\bibinfo
  {title} {Stability of the driving bunch in the plasma wakefield
  accelerator},\ }\href {https://doi.org/10.1109/TPS.1987.4316684} {\bibfield
  {journal} {\bibinfo  {journal} {IEEE Trans. Plasma Sci.}\ }\textbf {\bibinfo
  {volume} {15}},\ \bibinfo {pages} {192} (\bibinfo {year} {1987})}\BibitemShut
  {NoStop}%
\bibitem [{\citenamefont {Galyamin}\ \emph {et~al.}(2013)\citenamefont
  {Galyamin}, \citenamefont {Kapshtan},\ and\ \citenamefont
  {Tyukhtin}}]{Galyamin2013}%
  \BibitemOpen
  \bibfield  {author} {\bibinfo {author} {\bibfnamefont {S.~N.}\ \bibnamefont
  {Galyamin}}, \bibinfo {author} {\bibfnamefont {D.~Y.}\ \bibnamefont
  {Kapshtan}},\ and\ \bibinfo {author} {\bibfnamefont {A.~V.}\ \bibnamefont
  {Tyukhtin}},\ }\bibfield  {title} {\bibinfo {title} {Electromagnetic field of
  a charge moving in a cold magnetized plasma},\ }\href
  {https://doi.org/10.1103/PhysRevE.87.013109} {\bibfield  {journal} {\bibinfo
  {journal} {Phys. Rev. E}\ }\textbf {\bibinfo {volume} {87}},\ \bibinfo
  {pages} {013109} (\bibinfo {year} {2013})}\BibitemShut {NoStop}%
\bibitem [{\citenamefont {Galyamin}(2021)}]{Galyamin2021}%
  \BibitemOpen
  \bibfield  {author} {\bibinfo {author} {\bibfnamefont {S.~N.}\ \bibnamefont
  {Galyamin}},\ }\href@noop {} {\bibinfo {title} {Wakefields in hollow channel
  of magnetized plasma}} (\bibinfo {year} {2021}),\ \Eprint
  {https://arxiv.org/abs/2104.13828} {arXiv:2104.13828 [physics.plasm-ph]}
  \BibitemShut {NoStop}%
\bibitem [{\citenamefont {Molavi~Choobini}\ and\ \citenamefont
  {Shahmansouri}(2026)}]{MolaviChoobini2026}%
  \BibitemOpen
  \bibfield  {author} {\bibinfo {author} {\bibfnamefont {A.~A.}\ \bibnamefont
  {Molavi~Choobini}}\ and\ \bibinfo {author} {\bibfnamefont {M.}~\bibnamefont
  {Shahmansouri}},\ }\href@noop {} {\bibinfo {title} {Theoretical analysis and
  {PIC} simulations of electromagnetic wakefields excited by relativistic beams
  in magnetized plasmas}} (\bibinfo {year} {2026}),\ \Eprint
  {https://arxiv.org/abs/2604.25348} {arXiv:2604.25348} \BibitemShut {NoStop}%
\bibitem [{\citenamefont {Nersisyan}\ and\ \citenamefont
  {Elbakian}(2000)}]{Nersisyan2000}%
  \BibitemOpen
  \bibfield  {author} {\bibinfo {author} {\bibfnamefont {H.~B.}\ \bibnamefont
  {Nersisyan}}\ and\ \bibinfo {author} {\bibfnamefont {S.~S.}\ \bibnamefont
  {Elbakian}},\ }\bibfield  {title} {\bibinfo {title} {Excitation of nonlinear
  one-dimensional wake waves in underdense and overdense magnetized plasma by a
  relativistic electron bunch},\ }\href@noop {} {\bibfield  {journal} {\bibinfo
   {journal} {Part. Accel.}\ }\textbf {\bibinfo {volume} {63}},\ \bibinfo
  {pages} {279} (\bibinfo {year} {2000})},\ \Eprint
  {https://arxiv.org/abs/physics/9905046} {arXiv:physics/9905046} \BibitemShut
  {NoStop}%
\bibitem [{\citenamefont {Balakirev}\ \emph {et~al.}(2001)\citenamefont
  {Balakirev}, \citenamefont {Karas'}, \citenamefont {Karas'},\ and\
  \citenamefont {Levchenko}}]{Balakirev2001}%
  \BibitemOpen
  \bibfield  {author} {\bibinfo {author} {\bibfnamefont {V.~A.}\ \bibnamefont
  {Balakirev}}, \bibinfo {author} {\bibfnamefont {V.~I.}\ \bibnamefont
  {Karas'}}, \bibinfo {author} {\bibfnamefont {I.~V.}\ \bibnamefont {Karas'}},\
  and\ \bibinfo {author} {\bibfnamefont {V.~D.}\ \bibnamefont {Levchenko}},\
  }\bibfield  {title} {\bibinfo {title} {Plasma wake-field excitation by
  relativistic electron bunches and charged particle acceleration in the
  presence of external magnetic field},\ }\href
  {https://doi.org/10.1017/S0263034601194061} {\bibfield  {journal} {\bibinfo
  {journal} {Laser Part. Beams}\ }\textbf {\bibinfo {volume} {19}},\ \bibinfo
  {pages} {597} (\bibinfo {year} {2001})}\BibitemShut {NoStop}%
\bibitem [{\citenamefont {Karmakar}\ \emph {et~al.}(2017)\citenamefont
  {Karmakar}, \citenamefont {Chakrabarti},\ and\ \citenamefont
  {Sengupta}}]{Karmakar2017}%
  \BibitemOpen
  \bibfield  {author} {\bibinfo {author} {\bibfnamefont {M.}~\bibnamefont
  {Karmakar}}, \bibinfo {author} {\bibfnamefont {N.}~\bibnamefont
  {Chakrabarti}},\ and\ \bibinfo {author} {\bibfnamefont {S.}~\bibnamefont
  {Sengupta}},\ }\bibfield  {title} {\bibinfo {title} {Plasma wakefield
  excitation in a cold magnetized plasma for particle acceleration},\ }\href
  {https://doi.org/10.1063/1.4982808} {\bibfield  {journal} {\bibinfo
  {journal} {Phys. Plasmas}\ }\textbf {\bibinfo {volume} {24}},\ \bibinfo
  {pages} {052111} (\bibinfo {year} {2017})}\BibitemShut {NoStop}%
\bibitem [{\citenamefont {Katsouleas}\ and\ \citenamefont
  {Dawson}(1983)}]{Katsouleas1983}%
  \BibitemOpen
  \bibfield  {author} {\bibinfo {author} {\bibfnamefont {T.}~\bibnamefont
  {Katsouleas}}\ and\ \bibinfo {author} {\bibfnamefont {J.~M.}\ \bibnamefont
  {Dawson}},\ }\bibfield  {title} {\bibinfo {title} {Unlimited electron
  acceleration in laser-driven plasma waves},\ }\href
  {https://doi.org/10.1103/PhysRevLett.51.392} {\bibfield  {journal} {\bibinfo
  {journal} {Phys. Rev. Lett.}\ }\textbf {\bibinfo {volume} {51}},\ \bibinfo
  {pages} {392} (\bibinfo {year} {1983})}\BibitemShut {NoStop}%
\bibitem [{\citenamefont {Mofiz}(1989)}]{Mofiz1989}%
  \BibitemOpen
  \bibfield  {author} {\bibinfo {author} {\bibfnamefont {U.~A.}\ \bibnamefont
  {Mofiz}},\ }\bibfield  {title} {\bibinfo {title} {Wake-field accelerator in a
  magnetized electron-positron plasma},\ }\href
  {https://doi.org/10.1103/PhysRevA.40.6752} {\bibfield  {journal} {\bibinfo
  {journal} {Phys. Rev. A}\ }\textbf {\bibinfo {volume} {40}},\ \bibinfo
  {pages} {6752} (\bibinfo {year} {1989})}\BibitemShut {NoStop}%
\bibitem [{\citenamefont {Chen}\ \emph {et~al.}(2002)\citenamefont {Chen},
  \citenamefont {Tajima},\ and\ \citenamefont
  {Takahashi}}]{ChenTajimaTakahashi2002}%
  \BibitemOpen
  \bibfield  {author} {\bibinfo {author} {\bibfnamefont {P.}~\bibnamefont
  {Chen}}, \bibinfo {author} {\bibfnamefont {T.}~\bibnamefont {Tajima}},\ and\
  \bibinfo {author} {\bibfnamefont {Y.}~\bibnamefont {Takahashi}},\ }\bibfield
  {title} {\bibinfo {title} {Plasma wakefield acceleration for ultrahigh-energy
  cosmic rays},\ }\href {https://doi.org/10.1103/PhysRevLett.89.161101}
  {\bibfield  {journal} {\bibinfo  {journal} {Phys. Rev. Lett.}\ }\textbf
  {\bibinfo {volume} {89}},\ \bibinfo {pages} {161101} (\bibinfo {year}
  {2002})}\BibitemShut {NoStop}%
\bibitem [{\citenamefont {Chang}\ \emph {et~al.}(2009)\citenamefont {Chang},
  \citenamefont {Chen}, \citenamefont {Lin}, \citenamefont {Noble},\ and\
  \citenamefont {Sydora}}]{Chang2009}%
  \BibitemOpen
  \bibfield  {author} {\bibinfo {author} {\bibfnamefont {F.-Y.}\ \bibnamefont
  {Chang}}, \bibinfo {author} {\bibfnamefont {P.}~\bibnamefont {Chen}},
  \bibinfo {author} {\bibfnamefont {G.-L.}\ \bibnamefont {Lin}}, \bibinfo
  {author} {\bibfnamefont {R.}~\bibnamefont {Noble}},\ and\ \bibinfo {author}
  {\bibfnamefont {R.}~\bibnamefont {Sydora}},\ }\bibfield  {title} {\bibinfo
  {title} {Magnetowave induced plasma wakefield acceleration for ultrahigh
  energy cosmic rays},\ }\href {https://doi.org/10.1103/PhysRevLett.102.111101}
  {\bibfield  {journal} {\bibinfo  {journal} {Phys. Rev. Lett.}\ }\textbf
  {\bibinfo {volume} {102}},\ \bibinfo {pages} {111101} (\bibinfo {year}
  {2009})}\BibitemShut {NoStop}%
\bibitem [{\citenamefont {Bulanov}\ \emph {et~al.}(2013)\citenamefont
  {Bulanov}, \citenamefont {Esirkepov}, \citenamefont {Kando}, \citenamefont
  {Koga}, \citenamefont {Hosokai}, \citenamefont {Zhidkov},\ and\ \citenamefont
  {Kodama}}]{Bulanov2013}%
  \BibitemOpen
  \bibfield  {author} {\bibinfo {author} {\bibfnamefont {S.~V.}\ \bibnamefont
  {Bulanov}}, \bibinfo {author} {\bibfnamefont {T.~Z.}\ \bibnamefont
  {Esirkepov}}, \bibinfo {author} {\bibfnamefont {M.}~\bibnamefont {Kando}},
  \bibinfo {author} {\bibfnamefont {J.~K.}\ \bibnamefont {Koga}}, \bibinfo
  {author} {\bibfnamefont {T.}~\bibnamefont {Hosokai}}, \bibinfo {author}
  {\bibfnamefont {A.~G.}\ \bibnamefont {Zhidkov}},\ and\ \bibinfo {author}
  {\bibfnamefont {R.}~\bibnamefont {Kodama}},\ }\bibfield  {title} {\bibinfo
  {title} {Nonlinear plasma wave in magnetized plasmas},\ }\href@noop {}
  {\bibfield  {journal} {\bibinfo  {journal} {Phys. Plasmas}\ }\textbf
  {\bibinfo {volume} {20}},\ \bibinfo {pages} {083113} (\bibinfo {year}
  {2013})}\BibitemShut {NoStop}%
\bibitem [{\citenamefont {Rassou}\ \emph {et~al.}(2015)\citenamefont {Rassou},
  \citenamefont {Bourdier},\ and\ \citenamefont {Drouin}}]{Rassou2015}%
  \BibitemOpen
  \bibfield  {author} {\bibinfo {author} {\bibfnamefont {S.}~\bibnamefont
  {Rassou}}, \bibinfo {author} {\bibfnamefont {A.}~\bibnamefont {Bourdier}},\
  and\ \bibinfo {author} {\bibfnamefont {M.}~\bibnamefont {Drouin}},\
  }\bibfield  {title} {\bibinfo {title} {Influence of a strong longitudinal
  magnetic field on laser wakefield acceleration},\ }\href@noop {} {\bibfield
  {journal} {\bibinfo  {journal} {Phys. Plasmas}\ }\textbf {\bibinfo {volume}
  {22}},\ \bibinfo {pages} {073104} (\bibinfo {year} {2015})}\BibitemShut
  {NoStop}%
\bibitem [{\citenamefont {Zhao}\ \emph {et~al.}(2019)\citenamefont {Zhao},
  \citenamefont {Weng}, \citenamefont {Chen}, \citenamefont {Zeng},
  \citenamefont {Hidding}, \citenamefont {Jaroszynski}, \citenamefont
  {Assmann},\ and\ \citenamefont {Sheng}}]{Zhao2019}%
  \BibitemOpen
  \bibfield  {author} {\bibinfo {author} {\bibfnamefont {Q.}~\bibnamefont
  {Zhao}}, \bibinfo {author} {\bibfnamefont {S.~M.}\ \bibnamefont {Weng}},
  \bibinfo {author} {\bibfnamefont {M.}~\bibnamefont {Chen}}, \bibinfo {author}
  {\bibfnamefont {M.}~\bibnamefont {Zeng}}, \bibinfo {author} {\bibfnamefont
  {B.}~\bibnamefont {Hidding}}, \bibinfo {author} {\bibfnamefont {D.~A.}\
  \bibnamefont {Jaroszynski}}, \bibinfo {author} {\bibfnamefont
  {R.}~\bibnamefont {Assmann}},\ and\ \bibinfo {author} {\bibfnamefont {Z.~M.}\
  \bibnamefont {Sheng}},\ }\bibfield  {title} {\bibinfo {title}
  {Sub-femtosecond electron bunches in laser wakefield acceleration via
  injection suppression with a magnetic field},\ }\href@noop {} {\bibfield
  {journal} {\bibinfo  {journal} {Plasma Phys. Control. Fusion}\ }\textbf
  {\bibinfo {volume} {61}},\ \bibinfo {pages} {085015} (\bibinfo {year}
  {2019})}\BibitemShut {NoStop}%
\bibitem [{\citenamefont {Liu}\ \emph {et~al.}(2026)\citenamefont {Liu},
  \citenamefont {Chen}, \citenamefont {Lin}, \citenamefont {Gessner},\ and\
  \citenamefont {Hidding}}]{Letter}%
  \BibitemOpen
  \bibfield  {author} {\bibinfo {author} {\bibfnamefont {Y.-K.}\ \bibnamefont
  {Liu}}, \bibinfo {author} {\bibfnamefont {P.}~\bibnamefont {Chen}}, \bibinfo
  {author} {\bibfnamefont {C.-E.}\ \bibnamefont {Lin}}, \bibinfo {author}
  {\bibfnamefont {S.}~\bibnamefont {Gessner}},\ and\ \bibinfo {author}
  {\bibfnamefont {B.}~\bibnamefont {Hidding}},\ }\bibfield  {title} {\bibinfo
  {title} {Magnetizing nonlinear plasma wakefields for positron acceleration},\
  }\href@noop {} {\bibfield  {journal} {\bibinfo  {journal} {arXiv:2608.30455}\
  } (\bibinfo {year} {2026})},\ \bibinfo {note} {companion Letter, submitted to
  Phys.\ Rev.\ Lett.},\ \Eprint {https://arxiv.org/abs/2608.30455}
  {arXiv:2608.30455 [physics.acc-ph]} \BibitemShut {NoStop}%
\bibitem [{\citenamefont {Mora}\ and\ \citenamefont
  {Antonsen}(1997)}]{Mora1997}%
  \BibitemOpen
  \bibfield  {author} {\bibinfo {author} {\bibfnamefont {P.}~\bibnamefont
  {Mora}}\ and\ \bibinfo {author} {\bibfnamefont {T.~M.}\ \bibnamefont
  {Antonsen}, \bibfnamefont {Jr.}},\ }\bibfield  {title} {\bibinfo {title}
  {Kinetic modeling of intense, short laser pulses propagating in tenuous
  plasmas},\ }\href@noop {} {\bibfield  {journal} {\bibinfo  {journal} {Phys.
  Plasmas}\ }\textbf {\bibinfo {volume} {4}},\ \bibinfo {pages} {217} (\bibinfo
  {year} {1997})}\BibitemShut {NoStop}%
\bibitem [{\citenamefont {Busch}(1926)}]{Busch1926}%
  \BibitemOpen
  \bibfield  {author} {\bibinfo {author} {\bibfnamefont {H.}~\bibnamefont
  {Busch}},\ }\bibfield  {title} {\bibinfo {title} {Berechnung der {B}ahn von
  {K}athodenstrahlen im axialsymmetrischen elektromagnetischen {F}elde},\
  }\href@noop {} {\bibfield  {journal} {\bibinfo  {journal} {Ann. Phys.
  (Leipzig)}\ }\textbf {\bibinfo {volume} {386}},\ \bibinfo {pages} {974}
  (\bibinfo {year} {1926})}\BibitemShut {NoStop}%
\bibitem [{\citenamefont {Reiser}(2008)}]{Reiser2008}%
  \BibitemOpen
  \bibfield  {author} {\bibinfo {author} {\bibfnamefont {M.}~\bibnamefont
  {Reiser}},\ }\href@noop {} {\emph {\bibinfo {title} {Theory and Design of
  Charged Particle Beams}}},\ \bibinfo {edition} {2nd}\ ed.\ (\bibinfo
  {publisher} {Wiley-VCH},\ \bibinfo {address} {Weinheim},\ \bibinfo {year}
  {2008})\BibitemShut {NoStop}%
\bibitem [{\citenamefont {Derouillat}\ \emph {et~al.}(2018)\citenamefont
  {Derouillat}, \citenamefont {Beck}, \citenamefont {P{\'e}rez}, \citenamefont
  {Vinci}, \citenamefont {Chiaramello}, \citenamefont {Grassi}, \citenamefont
  {Fl{\'e}}, \citenamefont {Bouchard}, \citenamefont {Plotnikov}, \citenamefont
  {Aunai}, \citenamefont {Dargent}, \citenamefont {Riconda},\ and\
  \citenamefont {Grech}}]{Derouillat2018}%
  \BibitemOpen
  \bibfield  {author} {\bibinfo {author} {\bibfnamefont {J.}~\bibnamefont
  {Derouillat}}, \bibinfo {author} {\bibfnamefont {A.}~\bibnamefont {Beck}},
  \bibinfo {author} {\bibfnamefont {F.}~\bibnamefont {P{\'e}rez}}, \bibinfo
  {author} {\bibfnamefont {T.}~\bibnamefont {Vinci}}, \bibinfo {author}
  {\bibfnamefont {M.}~\bibnamefont {Chiaramello}}, \bibinfo {author}
  {\bibfnamefont {A.}~\bibnamefont {Grassi}}, \bibinfo {author} {\bibfnamefont
  {M.}~\bibnamefont {Fl{\'e}}}, \bibinfo {author} {\bibfnamefont
  {G.}~\bibnamefont {Bouchard}}, \bibinfo {author} {\bibfnamefont
  {I.}~\bibnamefont {Plotnikov}}, \bibinfo {author} {\bibfnamefont
  {N.}~\bibnamefont {Aunai}}, \bibinfo {author} {\bibfnamefont
  {J.}~\bibnamefont {Dargent}}, \bibinfo {author} {\bibfnamefont
  {C.}~\bibnamefont {Riconda}},\ and\ \bibinfo {author} {\bibfnamefont
  {M.}~\bibnamefont {Grech}},\ }\bibfield  {title} {\bibinfo {title} {Smilei: A
  collaborative, open-source, multi-purpose particle-in-cell code for plasma
  simulation},\ }\href@noop {} {\bibfield  {journal} {\bibinfo  {journal}
  {Comput. Phys. Commun.}\ }\textbf {\bibinfo {volume} {222}},\ \bibinfo
  {pages} {351} (\bibinfo {year} {2018})}\BibitemShut {NoStop}%
\bibitem [{\citenamefont {Vay}\ \emph {et~al.}(2018)\citenamefont {Vay},
  \citenamefont {Almgren}, \citenamefont {Bell}, \citenamefont {Ge},
  \citenamefont {Grote}, \citenamefont {Hogan}, \citenamefont {Kononenko},
  \citenamefont {Lehe}, \citenamefont {Myers}, \citenamefont {Ng},
  \citenamefont {Park}, \citenamefont {Ryne}, \citenamefont {Shapoval},
  \citenamefont {Th{\'e}venet},\ and\ \citenamefont {Zhang}}]{Vay2018}%
  \BibitemOpen
  \bibfield  {author} {\bibinfo {author} {\bibfnamefont {J.-L.}\ \bibnamefont
  {Vay}}, \bibinfo {author} {\bibfnamefont {A.}~\bibnamefont {Almgren}},
  \bibinfo {author} {\bibfnamefont {J.}~\bibnamefont {Bell}}, \bibinfo {author}
  {\bibfnamefont {L.}~\bibnamefont {Ge}}, \bibinfo {author} {\bibfnamefont
  {D.~P.}\ \bibnamefont {Grote}}, \bibinfo {author} {\bibfnamefont
  {M.}~\bibnamefont {Hogan}}, \bibinfo {author} {\bibfnamefont
  {O.}~\bibnamefont {Kononenko}}, \bibinfo {author} {\bibfnamefont
  {R.}~\bibnamefont {Lehe}}, \bibinfo {author} {\bibfnamefont {A.}~\bibnamefont
  {Myers}}, \bibinfo {author} {\bibfnamefont {C.}~\bibnamefont {Ng}}, \bibinfo
  {author} {\bibfnamefont {J.}~\bibnamefont {Park}}, \bibinfo {author}
  {\bibfnamefont {R.}~\bibnamefont {Ryne}}, \bibinfo {author} {\bibfnamefont
  {O.}~\bibnamefont {Shapoval}}, \bibinfo {author} {\bibfnamefont
  {M.}~\bibnamefont {Th{\'e}venet}},\ and\ \bibinfo {author} {\bibfnamefont
  {W.}~\bibnamefont {Zhang}},\ }\bibfield  {title} {\bibinfo {title} {Warp-{X}:
  A new exascale computing platform for beam-plasma simulations},\ }\href
  {https://doi.org/10.1016/j.nima.2018.01.035} {\bibfield  {journal} {\bibinfo
  {journal} {Nucl. Instrum. Methods Phys. Res. A}\ }\textbf {\bibinfo {volume}
  {909}},\ \bibinfo {pages} {476} (\bibinfo {year} {2018})}\BibitemShut
  {NoStop}%
\bibitem [{\citenamefont {Lifschitz}\ \emph {et~al.}(2009)\citenamefont
  {Lifschitz}, \citenamefont {Davoine}, \citenamefont {Lefebvre}, \citenamefont
  {Faure}, \citenamefont {Rechatin},\ and\ \citenamefont
  {Malka}}]{Lifschitz2009}%
  \BibitemOpen
  \bibfield  {author} {\bibinfo {author} {\bibfnamefont {A.~F.}\ \bibnamefont
  {Lifschitz}}, \bibinfo {author} {\bibfnamefont {X.}~\bibnamefont {Davoine}},
  \bibinfo {author} {\bibfnamefont {E.}~\bibnamefont {Lefebvre}}, \bibinfo
  {author} {\bibfnamefont {J.}~\bibnamefont {Faure}}, \bibinfo {author}
  {\bibfnamefont {C.}~\bibnamefont {Rechatin}},\ and\ \bibinfo {author}
  {\bibfnamefont {V.}~\bibnamefont {Malka}},\ }\bibfield  {title} {\bibinfo
  {title} {Particle-in-cell modelling of laser--plasma interaction using
  {F}ourier decomposition},\ }\href {https://doi.org/10.1016/j.jcp.2008.11.017}
  {\bibfield  {journal} {\bibinfo  {journal} {J. Comput. Phys.}\ }\textbf
  {\bibinfo {volume} {228}},\ \bibinfo {pages} {1803} (\bibinfo {year}
  {2009})}\BibitemShut {NoStop}%
\bibitem [{\citenamefont {Zemzemi}\ \emph {et~al.}(2020)\citenamefont
  {Zemzemi}, \citenamefont {Massimo},\ and\ \citenamefont
  {Beck}}]{Zemzemi2020}%
  \BibitemOpen
  \bibfield  {author} {\bibinfo {author} {\bibfnamefont {I.}~\bibnamefont
  {Zemzemi}}, \bibinfo {author} {\bibfnamefont {F.}~\bibnamefont {Massimo}},\
  and\ \bibinfo {author} {\bibfnamefont {A.}~\bibnamefont {Beck}},\ }\bibfield
  {title} {\bibinfo {title} {Azimuthal decomposition study of a realistic laser
  profile for efficient modeling of laser wakefield acceleration},\ }\href
  {https://doi.org/10.1088/1742-6596/1596/1/012054} {\bibfield  {journal}
  {\bibinfo  {journal} {J. Phys.: Conf. Ser.}\ }\textbf {\bibinfo {volume}
  {1596}},\ \bibinfo {pages} {012054} (\bibinfo {year} {2020})}\BibitemShut
  {NoStop}%
\bibitem [{\citenamefont {{The WarpX Development Team}}(2026)}]{WarpX2607}%
  \BibitemOpen
  \bibfield  {author} {\bibinfo {author} {\bibnamefont {{The WarpX Development
  Team}}},\ }\href {https://doi.org/10.5281/zenodo.21268496} {\bibinfo {title}
  {{WarpX}, version 26.07}} (\bibinfo {year} {2026}),\ \bibinfo {note} {rZ
  geometry, FDTD solver}\BibitemShut {NoStop}%
\bibitem [{\citenamefont {Hahn}\ \emph {et~al.}(2019)\citenamefont {Hahn},
  \citenamefont {Kim}, \citenamefont {Kim}, \citenamefont {Hu}, \citenamefont
  {Painter}, \citenamefont {Dixon}, \citenamefont {Kim}, \citenamefont
  {Bhattarai}, \citenamefont {Noguchi}, \citenamefont {Jaroszynski},\ and\
  \citenamefont {Larbalestier}}]{Hahn2019}%
  \BibitemOpen
  \bibfield  {author} {\bibinfo {author} {\bibfnamefont {S.}~\bibnamefont
  {Hahn}}, \bibinfo {author} {\bibfnamefont {K.}~\bibnamefont {Kim}}, \bibinfo
  {author} {\bibfnamefont {K.}~\bibnamefont {Kim}}, \bibinfo {author}
  {\bibfnamefont {X.}~\bibnamefont {Hu}}, \bibinfo {author} {\bibfnamefont
  {T.}~\bibnamefont {Painter}}, \bibinfo {author} {\bibfnamefont
  {I.}~\bibnamefont {Dixon}}, \bibinfo {author} {\bibfnamefont
  {S.}~\bibnamefont {Kim}}, \bibinfo {author} {\bibfnamefont {K.~R.}\
  \bibnamefont {Bhattarai}}, \bibinfo {author} {\bibfnamefont {S.}~\bibnamefont
  {Noguchi}}, \bibinfo {author} {\bibfnamefont {J.}~\bibnamefont
  {Jaroszynski}},\ and\ \bibinfo {author} {\bibfnamefont {D.~C.}\ \bibnamefont
  {Larbalestier}},\ }\bibfield  {title} {\bibinfo {title} {{45.5-tesla}
  direct-current magnetic field generated with a high-temperature
  superconducting magnet},\ }\href@noop {} {\bibfield  {journal} {\bibinfo
  {journal} {Nature}\ }\textbf {\bibinfo {volume} {570}},\ \bibinfo {pages}
  {496} (\bibinfo {year} {2019})}\BibitemShut {NoStop}%
\bibitem [{\citenamefont {Diederichs}\ \emph {et~al.}(2020)\citenamefont
  {Diederichs}, \citenamefont {Benedetti}, \citenamefont {Esarey},
  \citenamefont {Osterhoff},\ and\ \citenamefont {Schroeder}}]{Diederichs2020}%
  \BibitemOpen
  \bibfield  {author} {\bibinfo {author} {\bibfnamefont {S.}~\bibnamefont
  {Diederichs}}, \bibinfo {author} {\bibfnamefont {C.}~\bibnamefont
  {Benedetti}}, \bibinfo {author} {\bibfnamefont {E.}~\bibnamefont {Esarey}},
  \bibinfo {author} {\bibfnamefont {J.}~\bibnamefont {Osterhoff}},\ and\
  \bibinfo {author} {\bibfnamefont {C.~B.}\ \bibnamefont {Schroeder}},\
  }\bibfield  {title} {\bibinfo {title} {High-quality positron acceleration in
  beam-driven plasma accelerators},\ }\href
  {https://doi.org/10.1103/PhysRevAccelBeams.23.121301} {\bibfield  {journal}
  {\bibinfo  {journal} {Phys. Rev. Accel. Beams}\ }\textbf {\bibinfo {volume}
  {23}},\ \bibinfo {pages} {121301} (\bibinfo {year} {2020})}\BibitemShut
  {NoStop}%
\bibitem [{\citenamefont {Wang}\ \emph {et~al.}(2021)\citenamefont {Wang},
  \citenamefont {Khudik},\ and\ \citenamefont {Shvets}}]{Wang2021}%
  \BibitemOpen
  \bibfield  {author} {\bibinfo {author} {\bibfnamefont {T.}~\bibnamefont
  {Wang}}, \bibinfo {author} {\bibfnamefont {V.}~\bibnamefont {Khudik}},\ and\
  \bibinfo {author} {\bibfnamefont {G.}~\bibnamefont {Shvets}},\ }\href@noop {}
  {\bibinfo {title} {Positron acceleration in an elongated bubble regime}}
  (\bibinfo {year} {2021}),\ \Eprint {https://arxiv.org/abs/2110.10290}
  {arXiv:2110.10290 [physics.acc-ph]} \BibitemShut {NoStop}%
\end{thebibliography}%

\end{document}